\documentclass[authoryear,review]{elsarticle}
\usepackage{color}
\usepackage{amsmath,amssymb,graphicx}
\newcommand{\bC}{\mathbf{C}}

\newcommand{\bu}{\mathbf{u}}

\newcommand{\bw}{\mathbf{w}}

\newcommand{\E}{\mathbb{E}}

\newcommand{\be}{\begin{equation}}
\newcommand{\ee}{\end{equation}}
\newcommand{\ba}{\begin{equation} \begin{aligned}}
\newcommand{\ea}{\end{aligned} \end{equation}}

\begin{document}

\begin{frontmatter} 

\title{Temporally variable dispersal and demography can accelerate \\ the spread of invading 
species \tnoteref{t1}}
 
\tnotetext[t1]{\textbf{Last compile:} \today} 

\author[spe]{Stephen P. Ellner} \corref{cor1}
\ead{spe2@cornell.edu} 
 
\author[sjs]{Sebastian J. Schreiber}
\ead{sschreiber@ucdavis.edu}

\address[spe]{Department of Ecology and Evolutionary Biology, Cornell University,
Ithaca, NY 14853-2701, USA}

\address[sjs]{Department of Evolution and Ecology, and the Center for Population Biology \\
University of California, Davis, CA 95616, USA}

\cortext[cor1]{Corresponding author. Telephone 607-254-4221, FAX 607-255-8088.} 

\journal{Theoretical Population Biology} 

\begin{abstract}
We analyze how temporal variability in local demography and dispersal combine to affect
the rate of spread of an invading species. Our model combines state-structured local
demography (specified by an integral or matrix projection model) with general dispersal
distributions that may depend on the state of the individual or its parent. It 
allows very general patterns of stationary temporal variation in both local demography 
and in the frequency and distribution of dispersal distances. We show that expressions 
for the asymptotic spread rate and its sensitivity to parameters, 
which have been derived previously for less general models, continue to hold. Using
these results we show that random temporal variability in dispersal can accelerate population spread.
Demographic variability can further accelerate spread if it is positively correlated with
dispersal variability, for example if high-fecundity years are also years in which
juveniles tend to settle further away from their parents. A simple model for 
the growth and spread of patches of an invasive plant (perennial pepperweed, \textit{Lepidium latifolium})
illustrates these effects and shows that they can have substantial impacts on 
the predicted speed of an invasion wave. Temporal variability in dispersal has gotten
very little attention in both the theoretical and empirical literatures on invasive species spread. Our
results suggest that this needs to change.  
\end{abstract} 


\end{frontmatter} 

%\linenumbers
%\modulolinenumbers[2]

\section{Introduction}
\label{Intro}
Invasive organisms are ``altering the world's natural communities 
and their ecological character at an unprecedented rate'' \citep[p. 706]{mack-etal-00}, 
and often have substantial impacts on ecosystem structure and function \citep{simberloff-11}. 
Troublesome invasives in our current home states include zebra mussels, emerald ash borer, 
giant hogweed, avocado thrips, and smooth cordgrass (\textit{Spartina}). 

Spatial models of population spread have a potentially important role in 
evaluating and designing strategies for preventing or slowing the spread of
invasive species (e.g., \cite{taylor-hastings-04,grevsted-05,jongejans-etal-08,
bogich-etal-08,epanchin-hastings-10}). For well-studied species, simulation models 
allow detailed demographic modeling and accurate representations of 
landscape structure (e.g., \cite{higgins-etal-2000,
jongejans-etal-08,andrew-ustin-10,minor-gardner-11}). But simple ``strategic'' 
models (such as deterministic or stochastic matrix 
models) have often been useful for identifying the life stages or
demographic processes that are the best targets of opportunity for management efforts 
to preserve a native species or control an invasive (e.g., 
\cite{shea-kelley-98,heppell-etal-2000,morris-doak-02,lande-etal-03,shea-kelley-04,shea-etal-10}). 
Sensitivity analysis of the long-term population growth rate $\lambda$, or of the
long-term population spread rate $c^*$, have often been a key tool in these applications. 

In a seminal paper \citet{kot-etal-96} showed how integrodifference equations
could be used to model realistic patterns of organism redistribution (e.g., long-tailed
distributions rather than the Gaussian spread that results from classical reaction-diffusion
models), and gave a simple expression for the asymptotic rate of population spread.
They found that long-tailed dispersal distributions can give faster rates of spread than 
a Gaussian distribution with the same mean square displacement. This has been proposed as 
a resolution of ``Reid's Paradox'', the rapid northward advance of tree species after
the last glacial retreat \citep{clark-1998}. The analysis by \citet{kot-etal-96} 
was quickly extended to include temporal variability in local population
growth \citep{neubert-etal-00}, demographic stochasticity \citep{lewis-2000},
discrete stage structure \citep{neubert-caswell-00}, and two-dimensional
spread \citep{lewis-etal-06}. Two recent extensions have been models 
with continuous population structure \citep{jongejans-etal-11} 
and models that combine discrete stage structure and demographic variability 
\citep{schreiber-ryan-11,caswell-etal-11}. \citet{caswell-etal-11} also provide
formulas for sensitivity analysis of the long-term population spread rate for
periodic or stochastic environmental variation. 

Here we take two additional steps. The first is the natural step of combining 
demographic variability with continuous population structure, using
the formalism of integral projection models. The second is to analyze the
effects of an ecologically important aspect of population spread that has 
received surprisingly little attention 
in previous work: temporal variability in dispersal.  Empirical evidence
is very limited, but suggests that the frequency and range of long-distance
dispersal can vary greatly from one year to the next \citep{andrew-ustin-10}. 
The general formulas for population spread rate and its sensitivity
in models with temporal variability \citep{neubert-etal-00, 
schreiber-ryan-11,caswell-etal-11} allow for temporal variability in both 
local demography and dispersal, but previous to this paper there
has not been (to our knowledge) a mathematical analysis of how dispersal variability
can affect the rate of population spread. However, a recent numerical study by
\citet{seo-lutscher-11} did examine periodic fluctuations in dispersal rates for  populations with sedentary and randomly diffusing individuals. They found that fluctuations in  dispersal rates could increase or decrease rates of spatial spread depending on temporal correlations between dispersal and demography. To better understand these and other interactions between demography and dispersal on population spread, we analyze three forms of variability, separately and in combination: temporal variation in local demographic parameters (e.g. survivorship and fertility), a single mode of dispersal whose parameters (e.g., mean dispersal distance) vary over time, and multiple modes of dispersal (e.g., local wind dispersal and long-range animal dispersal) whose frequencies vary over time.

Our analysis reveals that the effects of dispersal variability can be
very different from those of demographic variability. A classical 
\citep{lewontin-cohen-69} and very general result \citep{tuljapurkar-90,rees-ellner-09}
is that temporally uncorrelated demographic variability reduces population growth
and spread rates \citep{lewis-2000, clark-etal-2001,schreiber-ryan-11,
caswell-etal-11}. In contrast, we find that temporally uncorrelated dispersal variability can increase the 
rate of population spread. Moreover, when dispersal is variable rather
than constant, the effect of demographic variability can be reversed:
demographic variability that by itself would decrease population growth
and spread rate can instead increase those rates, if it is 
correlated with dispersal variability. These general results are
all derived by perturbation analysis for small fluctuations, but we
also provide a simple geometric explanation for the effect of dispersal
variability. We then use an empirically-based model for the spread of
an invasive plant (perennial pepperweed) to show that our results continue to hold at
very high levels of variability and that dispersal variability and
dispersal-demography covariance can have appreciable effects on population
spread rate. 

\section{Model and assumptions}
\label{LinearModel}
We consider a continuously structured population in which the state $z$ of an 
individual (e.g. size or age) lies in a compact set of all possible 
individual states $\mathbf{Z}$  \citep{ellner-rees-06}. These individuals disperse 
along a one dimensional transect of their environment (however, rates of spread in a two 
dimensional region can be computed by ``marginalizing'' 
a two-dimensional dispersal kernel along the direction of interest \citep{lewis-etal-06}). 
Consequently, the location $x$ of an individual can be 
identified with a point on the line $\mathbf{X}=(-\infty,\infty)$.  Let $n_t(x,z)$ denote the 
population density at location $x$, state $z$, and time $t$. In the absence of density dependence
(which we will consider in section \ref{NonLinModels}), the most general form of the model is 
\begin{equation}
n_{t+1}(x,z) = \iint {K_t(x,z,x_0,z_0)n_t(x_0 ,z_0)dx_0 dz_0 }
\label{eqn:fullmodel}
\end{equation}
where $x$ is location, $z$ is individual state, $n_t(x,z)$ is the population distribution in
space and state at time $t$, $K_t$ is the kernel for year $t$, and the integral runs over the spatial 
domain $\mathbf{X}=(-\infty,\infty)$ and the (compact) set of possible individual states
$\mathbf{Z}$. 
The kernel $K_t(x,z,x_0,z_0)$ represents the rate at which individuals in state $z_0$ and 
location $x_0$ at time $t$ produce individuals in state $z$ and location $x$ at time $t$. 
It includes changes in individual state, changes in location, and production 
of new offspring which may vary in state and location. We will often write
$n_{t+1}=K_t n_t$ as a shorthand for equation \eqref{eqn:fullmodel}, and similarly
for other kernels. 

Consistent with prior studies~\citep{kot-etal-96,neubert-caswell-00,neubert-etal-00,
jongejans-etal-11,caswell-etal-11,schreiber-ryan-11}, we assume spatial homogeneity. In particular, 
state transition rates are the same at all locations, so our model incorporates temporal variability
in environmental conditions but not spatiotemportal variability, and movement probability 
is a function of the distance between the starting and ending locations. That is,
\begin{equation}
K_t(x,z,x_0,z_0) = K_t(x-x_0,z,z_0), \mbox{ with } K_t(v,z,z_0)=K_t(-v,z,z_0).
\label{eqn:homog}
\end{equation}
Within that constraint, however, the dispersal pattern can depend on individual state 
in any way, in principle. Any constraints on movement dictated by the species' life history 
is reflected in the structure of the kernel. For example, if new offspring undergo natal 
dispersal (e.g., seeds or larvae) but then settle for the rest of their life 
(e.g., trees, corals), the kernel has the form
\begin{equation}
K_t(v,z,z_0)=\delta_0(v)P_t(z,z_0)+k_{d,t}(v)F_t(z,z_0)
\label{eqn:PlantKernel}
\end{equation}
where $\delta_0$ is the Dirac delta-function (a unit mass at $v=0$), $F$ and $P$ are the fecundity and 
survival/growth kernels respectively, and $k_{d,t}$ is the juvenile dispersal kernel that describes
the displacements of offspring from their parent. Without loss of generality we assume that $k_{d,t}$ 
is a probability distribution, i.e., that any offspring mortality prior to establishment is absorved
into $F$.  

To ensure that invasion speeds are well-defined, we need several additional assumptions. 
First, we assume that dispersal events have exponentially bounded tails. More precisely, 
we assume that the transformed kernels
\begin{equation}
H_{s,t}(z,z_0) = \int {K_t\left( {v,{z},{z_0}} \right){e^{sv}}dv}
\label{eqn:skernels}
\end{equation}
are finite with probability 1 for all $s$ in some interval $(-s_1,s_1)$; the interval is symmetric 
because of our spatial homogeneity assumption. In the case of juvenile dispersal, 
equation \eqref{eqn:PlantKernel}, the transformed kernels are  
\begin{equation}
H_{s,t} = P_t + M_t(s)F_t
\label{eqn:PlantH}
\end{equation}
where $M_t$ is the moment-generating function of $k_{d,t}$. In equations \eqref{eqn:skernels}
and \eqref{eqn:PlantH}, $s$ characterizes the shape of the invasion wave, and the kernels $H_{s,t}$
determine the spread rate for an invasion wave where total population density
decreases exponentially at rate $s$ as a function of distance from the population center
(\ref{Appendix:WaveSpeed}). Without an exponentially bounded tail, the rate of spatial spread may constantly 
accelerate, so there is no asymptotic invasion speed~\citep{kot-etal-96}.

Second, we assume that for all $s$ in $(-s_1,s_1)$, the temporal sequence of transformed kernels
$\{ \dots H_{s,-1},H_{s,0},H_{s,1},\dots \}$ are stationary, ergodic and satisfy the assumptions 
of \citet{ellner-rees-07} for stochastic integral projection models. Stationarity means that
the pattern of temporal variability in the environment doesn't change over time: each year
is (randomly) different from the previous, but there are no long-term trends in the statistical
properties of the variability, such as an increase in rainfall or temperature. 
The other main assumptions are that the kernels must be continuous, bounded,
and in an \textit{ergodic set}. The ergodic set property is analogous to the requirement that a 
matrix projection model be irreducible and aperiodic. It means that for some $m>0$, 
$H_{s,m}\cdots H_{s,2} H_{s,1}$ is positive at all $(z,z_0)$ with probability one, where 
the kernel product is defined as
$$H_2 H_1(z,z_0) = \int {H_2(z,y)H_1(y,z_0) dy} $$
representing the effect of $H_1$ acting on the population followed by $H_2$.
For the juvenile dispersal kernel \eqref{eqn:PlantKernel}, the transformed kernels will 
satisfy this assumption if the base kernels do and the temporal variability in 
$M_t(s)$ is bounded ($-\infty < m_1 \le M_t(s) \le m_2 < +\infty$ with probability 1). 

Under these assumptions, the dominant Lyapunov exponent $\gamma(s)$ of the
random sequence of transformed kernels exists, i.e. there is a number $\gamma(s)>0$ such that
\begin{equation}
\lim_{t\to\infty} \frac{1}{t} \log \| H_{s,t} H_{s,t-1} \dots H_{s,1}\| = \gamma(s)  
\end{equation}
with probability one. Here, $\|H_{s,t}\|$ denotes any operator norm of $H_{s,t}$. 

\section{Population growth and spatial spread in the linear model}
\label{LinearWave}
We begin by examining population growth and spatial spread in the absence of density-dependent feedbacks.
This analysis provides key information about how  the population grows and spreads when rare. In particular, when
the population experiences negative density-dependence, it determines the rate of spatial spread
at the leading edge of the invasion and hence the asymptotic invasion speed (see section \ref{NonLinModels}).

To understand asymptotic population growth without density-dependent feedbacks, we integrate both sides of
\eqref{eqn:fullmodel} over space and arrive at the overall state distribution $\tilde n_t(z)$ defined as  
$$\tilde n_t(z) = \int{n_t(x,z)dx}$$
So long as it is finite, this overall state distribution changes according to the
demographic kernel $ \tilde K_t(z,z_0) = H_{0,t}(z,z_0)=\int{K_t(v,z,z_0)dv}$, i.e.
\begin{equation}
\tilde n_{t+1}(z)= \int {\tilde K_t(z,z_0)\tilde n_t(z_0)dz_0 }.
\label{eqn:demogmodel}
\end{equation}
The total population summed over all locations is therefore described by a nonspatial stochastic
model. So starting from a finite initial population, the total abundance exhibits
asymptotically exponential growth or decay. Whether the population grows or shrinks is determined by the
dominant Lyapunov exponent $\gamma(0)$ of the the random kernels $\tilde K_t$. In the ecological literature the
dominant Lyapunov exponent is usually called the ``stochastic growth rate'' $\log \lambda_S$ where $\lambda_S$ 
is analogous to the dominant eigenvalue of a matrix projection model. Under our
assumptions, if the population starts from a finite initial distribution, then
$\lim_{t\to\infty} \frac{1}{t} \ln \int \int n_t(x,z)dxdz= \log \lambda_S$ with probability one. 
So if $\lambda_S >1$ (equivalently, if $\gamma(0)>0$) the
population grows exponentially, and if $\lambda_S <1$ it shrinks exponentially, in the long run.
If the environmental variability is not too strongly autocorrelated, then the probability distribution
of total population size is approximately lognormal, and the mean and variance of
log-transformed total population size are both linear functions of time~\citep{ellner-rees-07}. 

When $\lambda_S>1$, the total population size increases and one would expect the 
total area occupied by the population to increase. In an
infinite-population model such as ours, ``occupied'' has to be defined by a threshold density $n_c$,  
so that location $x$ is regarded as occupied at time $t$ if the total population (or total
biomass, or some other measure of total population size) at location $x$ at time $t$ is at least $n_c$. When the initial
spatial and state distribution of the population is of the form $n(x,z)=n_0(z)e^{-sx}$ with $s>0$, we show 
in \ref{Appendix:WaveSpeed} that the asymptotic wave speed is
\be
\label{eqn:cs}
c(s)=\frac{\gamma(s)}{s} 
\ee
where $\gamma(s)$ is the dominant Lyapunov exponent of the transformed kernels $H_{s,t}$. But unlike these initial 
distributions, real invasions begin with populations whose densities are bounded and restricted to a finite spatial interval. For these realistic invasions, we show in \ref{Appendix:WaveSpeed} that the 
asymptotic wave speed is bounded above by   
\be
c^* = \mathop {\min }\limits_{s > 0} c(s) .
\label{eqn:cstar}
\ee
Therefore, $c^*$ is the only biologically relevant wave speed. 

The relevance of $c^*$ hinges on the conjecture
that an initially localized invasion will asymptotically spread as a wave, and therefore at rate $c^*$.
Simulations for unstructured and matrix population models universally support this conjecture
for dispersal kernels with tails that decrease exponentially or faster,
so that a finite moment generating function exists. However, this has been proved rigorously
only for unstructured models under some additional assumptions about the shape of the dispersal kernel
(see Mollison 1991).   

Our result generalizes prior results on wave speeds in fluctuating environments. 
\citet{neubert-etal-00} found the same formula for unstructured populations living in a temporally 
uncorrelated environment. More recently, \citet{caswell-etal-11} and \citet{schreiber-ryan-11} 
independently derived this formula for matrix models (i.e. the kernel $K$ is supported by a 
finite number of states) in random environments. \citet{schreiber-ryan-11} also showed that the average
spread rate up to time $t$ is normally distributed with mean $c^*$ and a variance that decreases  
inversely proportional to $t$. The results of \citet{ellner-rees-07} imply that the same conclusion 
applies to the integral projection models considered here, when the environmental variability is
not too strongly autocorrelated. 

\section{Density-dependent models} \label{NonLinModels}
To account for density-dependent effects on dispersal and demography, we allow the transition kernel for individuals at
location $x_0$ to depend on the local population density $n(x_0)$. Then the most general form of the model is
\begin{equation}
n_{t+1}(x,z) = \int {K_t(n_t(x_0);x,z,x_0,z_0)n_t(x_0 ,z_0)dx_0 dz_0 }
\label{eqn:nonlinear}
\end{equation}
where the kernel $K_t(n_t(x_0);x,z,x_0,z_0)$ depends on the local density at $x_0$. Despite this additional 
complexity, there is compelling evidence that the asymptotic invasion speed is determined
by the linearization at $n=0$. This broad principle is called the ``linearization conjecture'' and is 
expected to hold provided per-capita survivorship and reproduction are greatest at low densities~\citep{mollison-91}. 
This conjecture has been mathematically verified for structured populations exhibiting compensating 
density dependence and living in constant or periodic environments~\citep{liu-89,weinberger-02}. Moreover, 
there is growing numerical support for this conjecture for structured  populations 
exhibiting overcompensating density-dependence and living in constant or fluctuating
environments~\citep{neubert-caswell-00,neubert-etal-00,schreiber-ryan-11,caswell-etal-11}.

For the general nonlinear model \eqref{eqn:nonlinear} considered here, the linearization conjecture applies provided two assumptions are met. First, population growth and dispersal is greatest at low densities i.e. $K_t(0;x,z,x_0,z_0)\ge K_t(n_t(x_0);x,z,x_0,z_0)$. Second, the kernel for the unoccupied habitat $K_t(0;x,z,x_0,z_0)$ satisfies the assumptions discussed in the previous section i.e. spatial homogeneity $K_t(0;x_0+v,z,x_0,z_0)=K_t(0; v, z, z_0)$, and  the existence, stationarity and ergodicity of the transformed kernels $H_{s,t}(z,z_0)=\int K_t(0;v,z,z_0) e^{-sv}dv$. The first assumption is violated for populations exhibiting positive density dependence at low densities (an Allee effect). For such populations, whether the population spreads or not can depend on the initial size and spatial distribution of the population~\citep{kot-etal-96}. When spatial spread occurs, these populations grow and spread faster when they achieve higher densities. Consequently, the linearization conjecture can underestimate asymptotic invasion speeds for populations exhibiting an Allee effect.

\section{Variable dispersal and spread rate}
A consistent finding in integrodifference models of population spread has been that temporally uncorrelated 
demographic variability slows the rate of spread~\citep{neubert-etal-00,caswell-etal-11,schreiber-ryan-11}. 
The root of this phenomenon is the small-fluctuations approximation for the stochastic growth 
rate $\gamma = \log \lambda_S$, see, for example, \eqref{eqn:SmallVar}. To leading order, any 
kind of temporally uncorrelated variability about average demographic rates causes the population 
growth rate to decrease in proportion to the interannual 
variance~\citep{lewontin-cohen-69,tuljapurkar-90,ellner-rees-07}.

In contrast, as we will now show, temporal variability in the range of dispersal can increase the 
rate of spread. A heuristic explanation for this difference is that 
Jensen's inequality acts in opposite directions on demographic and dispersal variation. For 
demographic variation, population growth depends (roughly) on the geometric mean of annual 
population growth rates, or equivalently on the mean of log-transformed annual growth rates. 
The logarithm function has negative second derivative, so variability decreases the mean. 
Dispersal variation affects the spread rate through $M(s)$, the moment generating function 
of the dispersal distribution. $M(s)$ has a positive second derivative (see \ref{Appendix:mgf}), 
so variation in dispersal can increase the mean of the kernels $H_{s,t}$ that determine the 
spread rate. In many situations the increase in mean dominates the negative impact of the 
increased variability due to dispersal variation, so spread rate increases.

This is the most technical section of the paper. Because our main conclusions have
been stated in the preceding paragraph and the Abstract, readers who wish can avoid the technical
details by skipping from here to section \ref{sect:DoesItMatter}. 

\subsection{Sensitivity formulas for invasion speed}
Our results are based on analyzing the effects of small perturbations to the kernel $K_{s,t}$. 
A small (order $\varepsilon$) perturbation of $K$ causes a perturbation to the 
transformed kernels from $H_{s,t}$ to $H_{s,t}+\varepsilon\, C_{s,t}$ for some
generally random kernel $C$. To approximate the effects of this perturbation using Taylor series, we need
derivatives with respect to $\varepsilon$ at $\varepsilon=0$. 

Two general results which apply to any perturbation are 
\be
\label{eqn:s0}
\frac{\partial^n c^*}{\partial \varepsilon^n} = 
\left[\frac{1}{s}{\frac{\partial^n \gamma}{\partial \varepsilon^n}}\right]_{s=s^*} 
\ee
\be
\label{eqn:s1}
\frac{\partial c^*}{\partial \varepsilon} = 
\frac{1}{s^*} \E\left[ {\frac{{\left\langle {{v_{s^*,t + 1}},{C_{s^*,t}}{w_{s^*,t}}} \right\rangle }}
{{\left\langle {{v_{s^*,t + 1}},{H_{s^*,t}}{w_{s^*,t}}} \right\rangle }}} \right]
\ee
where $\E$ denotes the expectation of a random variable, $\langle f, g \rangle= \int f(x)\cdot g(x) \,dx$ denotes the inner product of two integrable functions, the invasion speed $c^*$ associated with the base kernel $H_{s,t}$ occurs as $s=s^*$, and $v_{s,t},w_{s,t}$ 
are the time-dependent stationary reproductive value and population structure sequences 
of the base kernel $H_{s,t}$. Some important perturbations (such as adding variability in dispersal distances) affect the ``noise'' 
structure but have no mean effect, i.e., $E[C_{s,t}]=0$. Provided that 
the $H_{s,t}$ and $C_{s,t}$ are independent and identically distributed, and $C_{s,t}$ is independent of $H_{s,t}$, the 
effect of these random perturbations on invasion speed is of order $\varepsilon^2$: 
\be
\label{eq:s2}
\frac{\partial c^*}{\partial \varepsilon} =0 \mbox{ and } \frac{\partial^2 c^*}{\partial \varepsilon^2}
=- \frac{1}{s^*}\E\left[ {\frac{{{{\left\langle {{v_{s^*,t + 1}},{C_{s^*,t}}{w_{s^*,t}}} \right\rangle }^2}}}
{{{{\left\langle {{v_{s^*,t + 1}},{H_{s^*,t}}{w_{s^*,t}}} \right\rangle }^2}}}} \right].
\ee
The effect of the perturbation on asymptotic wave speed is therefore $\varepsilon^2/2$ times the
right-hand side of \eqref{eq:s2}. Equation \eqref{eq:s2} also holds if  $C_{s,t}$ is not independent
of the unperturbed process but its mean, conditional on any event in the unperturbed process, is
identically zero.   

The derivations of \eqref{eqn:s0}, \eqref{eqn:s1} and \eqref{eq:s2} are 
in \ref{Appendix:sensitivity}. Methods for numerically evaluating the expectations in \eqref{eqn:s1} and
\eqref{eq:s2} are given by \cite{rees-ellner-09}.

\subsection{Single dispersal mode}
As a simple illustration of how variable dispersal can accelerate spread, 
consider the juvenile dispersal model \eqref{eqn:PlantKernel}
when the juvenile dispersal kernel has constant shape but varying mean dispersal
distance. That is, $k_{d,t}(v) = (1/L_t)k_1(v/L_t)$ where $k_1$ is a dispersal kernel 
satisfying our assumptions in which the mean absolute parent-offspring displacement is one unit of distance. 
The scaling by $L_t$ guarantees that $k_{d,t}$ is a probability distribution, and 
$L_t$ is then the mean absolute parent-offspring displacement in year $t$.
Let $M_1(s)$ be the moment-generating function of $k_1$; then
\begin{equation}
H_{s,t} = P_t+ M_1(L_t s)F_t.
\end{equation}
For small fluctuations in $L_t$, we let $L_t = \bar L + \sigma_L z_t$
where $\E[z_t]=0, \mathrm{Var}[z_t]=\E[z_t^2]=1$ and $\sigma_L \ll 1$. We assume that the $z_t$ are independent 
and identically distributed, and independent of the demographic kernels $P_t,F_t$. Using the 
small-variance approximation
\be
M_1(L_t s) = M_1(\bar L s + s \sigma_L z_t) \approx M_1(\bar L s) + s\sigma_L M_1'(\bar L s)z_t
+ \frac{s^2\sigma_L^2 M_1''(\bar L s)}{2} z_t^2,
\label{eqn:SmallVarM}
\ee
the transformed kernel is
\be
H_{s,t} \approx  \underbrace{P_t +  M_1(\bar L s)F_t}_{H_{s,t}^0} + s\sigma_L M_1'(\bar L s) z_t F_t 
+ \frac{s^2 \sigma_L^2 M_1''(\bar L s)}{2}z_t^2F_t.
\label{eqn:SmallVarH}
\ee
The first two terms on the right-hand side of \eqref{eqn:SmallVarH} are the constant-dispersal 
transformed kernel which we denote $H^0_{s,t}$, and the last two terms are the 
perturbation due to variable dispersal.

The last term on the right-hand side of \eqref{eqn:SmallVarH} has positive mean.  
Using the sensitivity formula~\eqref{eqn:s1} and the independence of $z_t$, the
increase in the mean of the transformed kernel due to this term
increases the invasion speed by (see  \ref{TwoParameter} for details) 
\be
\frac{s^* \sigma_L^2 M_1''(\bar L s^*)}{2}
\E\left[ \frac{\left\langle v_{s^*,t}, F_t w_{s^*,t} \right\rangle}{\left\langle v_{s^*,t}, H^0_{s^*,t} w_{s^*,t} \right\rangle} \right]
\label{eqn:z2term}
\ee
where $c(s^*)=c^*$ is the invasion speed for the unperturbed kernel $H_{s,t}^0$, and $v_{s^*,t},w_{s^*,t}$ are the 
time-dependent stationary reproductive value and state distribution for $H_{s^*,t}^0$. 
Opposing this effect is a decrease in the invasion speed caused by the kernel variance. 
The dominant contribution to the kernel variance
comes from the $z_t$ term on the right-hand side of \eqref{eqn:SmallVarH}, which has zero mean. Using the perturbation
formula \eqref{eq:s2} with $\sigma_L$ as the small parameter $\varepsilon$, this term decreases the invasion speed by
\be
\frac{s^* \sigma_L^2 (M_1'(\bar L s^*))^2} {2}
\E\left[ \frac{\left\langle v_{s^*,t}, F_t w_{s^*,t} \right\rangle^2}{\left\langle v_{s^*,t}, H^0_{s^*,t} w_{s^*,t} \right\rangle^2} \right].
\label{eqn:zterm}
\ee
Variable dispersal will increase the spread rate whenever \eqref{eqn:z2term} is greater than \eqref{eqn:zterm}.
Because $H_{s,t}^0 \ge M_1(\bar L s) F_t$, we have
\be
\frac{\left\langle v_{s,t}, F_t w_{s,t} \right\rangle^2}{\left\langle v_{s,t}, H^0_{s,t} w_{s,t} \right\rangle^2}
\le \frac{1}{M_1(\bar L s)} \frac{\left\langle v_{s,t}, F_t w_{s,t} \right\rangle}{\left\langle v_{s,t}, H^0_{s,t} w_{s,t} \right\rangle}.
\label{eqn:vbound}
\ee
Therefore \eqref{eqn:z2term} is greater than \eqref{eqn:zterm} if $M_1'' M_1 > (M_1')^2.$
We show in \ref{Appendix:mgf} that this condition always holds, at all points where the moment generating function
is finite. Thus, small variation in the range of dispersal (added to variability in survival, growth and fecundity)
always increases the asymptotic rate of spread in the juvenile dispersal model.  

In \ref{StateDependent} we generalize this result to state-dependent dispersal, i.e. a 
dispersal kernels $k_{d,t}(v) = (1/L_t(z,z_0))k_1(v/L_t(z,z_0);z,z_0)$ so that movement distributions
can depend on both the parent and offspring states. For example, taller plants would be expected to 
have longer average seed dispersal distances. If the temporal variance in $L_t$ is small, and 
in any one year $L_t$ is either above-average for all $(z,z_0)$ or below-average 
for all $(z,z_0)$, then variance in $L_t$ increases the asymptotic spread rate. 

While this simple example illustrates that temporal variation in mean dispersal distances can increase the 
asymptotic invasion speed, not all forms of dispersal variation increase invasion speeds.  
Moreover, there can be subtle, nonlinear interactions between forms of dispersal variation that 
individually have opposing effects on invasion speeds. To illustrate these opposing trends and nonlinear
interactions, we examine next populations with multiple dispersal modes. 

\subsection{Multiple dispersal modes}
Organism frequently use different modes of dispersal. These modes may corresponds to different 
environmental currents (e.g., marine propagules at different heights in the water column experiencing 
different water currents), multiple vectors (e.g. plants dispersed by bird, mammal or insects), 
or individual differences in physiology or behavior. Let $k_i(v)$ be the dispersal kernel 
associated with the $i$-th mode of dispersal and $p_i$ the probability of an individual utilizing 
this mode of dispersal. Then the population level dispersal kernel equals $k(v)=\sum_{i=1}^n p_i k_i (v)$ 
and the associated moment generating function is $M(s)=\sum_{i=1}^n p_i m_i(s)$ where $m_i$ is the 
moment generating function for $k_i$.

We consider two forms of fluctuations on these multiple modes of dispersal:
small fluctuations in the frequencies $p(t)=(p_1(t),\dots,p_n(t))$ of these different modes of dispersal, 
and small fluctuations in the mean dispersal distance $L_i(t)$ for each of the modes of dispersal. 
For simplicity, we focus on the case where there are no fluctuations in local demography i.e.  $F_t=F$ and $P_t=P$ for all $t$.  The dispersal-demography kernel, in this case, equals  $H_{s,t}=P+ \sum_i p_i(t) m_i( L_i(t) s) F$.
Define 
\begin{equation*}
\begin{aligned}
\ell_i(t)&= L_i(t)-\bar L_i, \quad \varrho_i(t)=p_i(t)-\bar p_i, \\
m_i&=m_i(\bar L_i s^*), \quad m_{i,1}=m_i'(\bar L_i s^*) s^*, \quad m_{i,2}=m_i''(\bar L_i s) (s^*)^2/2
\end{aligned}
\end{equation*}
where $\bar L_i= \E[L_i(t)]$ is the expected mean dispersal distance for dispersal mode $i$, $\bar p_i$ is the expected frequency of dispersal mode $i$, and $s^*$ determines the invasion speed $c^*=c^*(s^*)$ for the unperturbed kerned $H_s^0 = P+ \sum_i \bar p_i m_i(\bar L_i s) F$.  If we assume the fluctuations are small (of order $\varepsilon$), a Taylor expansion to second order of the kernel $H_{s,t}$ at $s=s^*$  yields
\begin{eqnarray*}
H_{s^*,t}&\approx &   P+ \sum_i p_i(t) (m_i+m_{i,1}\ell_i(t)+m_{i,2}\ell_i(t)^2) F\\
&\approx& \underbrace{P+ \sum_i \bar p_i m_i F}_{H^0_{s^*}} + \underbrace{\sum_i\left( \varrho_i(t) m_i +\bar p_i m_{i,1}\ell_i(t) \right)F}_{A_t} 
+\underbrace{\sum_i ( \varrho_i(t) m_{i,1} \ell_i(t) + \bar p_i m_{i,2} \ell_i(t)^2) F}_{B_t}.
\end{eqnarray*}
 The perturbation $A_t$ has mean zero and variance of order $\varepsilon^2$. The perturbation $B_t$ has non-zero mean of order $\varepsilon^2$ and variance of order $\varepsilon^4$ and, consequently, this variance is negligible.
 
Let $\lambda_0$, $v$, and $w$ be the the dominant eigenvalue, the reproductive value and the stable state distribution for $H_{s^*}^0$, respectively. Using the sensitivity formula \eqref{eqn:s1} and the fact 
that $\langle v ,H^0_{s^*} w\rangle = \lambda_0$, the correction term to the invasion speed $c^*$ due to the perturbation $B_{t}$  is 
\begin{equation}
\frac{1}{s^*\lambda_0}\E[\langle v, B_{t} w \rangle] = \frac{\langle v, Fw \rangle}{s^*\lambda_0} \sum_i  \left(m_{i,1}\mbox{Cov}[\varrho_i(t),\ell_i(t)]+ \bar p_im_{i,2} \mbox{Var}[\ell_i(t)]\right).
\label{Bterm-1}
\end{equation}  
Since the perturbation $A_t$ has zero mean, the sensitivity formula \eqref{eq:s2} implies that the correction term to the invasion speed $c^*$ due to  the perturbation $A_{t}$  equals
\begin{equation}
\begin{aligned}
 - \dfrac{1}{2 s^* \lambda_0^2} \E [ \langle v, A_t w \rangle^2 ]
& = - \dfrac{\langle v, Fw \rangle^2}{2 s^* \lambda_0^2}\mbox{Var} \left[ \sum_i \varrho_i(t)m_i+\bar p_i m_{i,1}\ell_i(t)\right]. 
\label{Aterm-1}
\end{aligned} 
\end{equation} 
We can expand the variance term as 
\begin{equation}\label{eq:modes3b}
\begin{aligned}
\mbox{Var} \left[ \sum_i \varrho_i(t)m_i+\bar p_i m_{i,1}\ell_i(t)\right] =&\mbox{Var} \left[ \sum_i \varrho_i(t)m_i\right] +\mbox{Var} \left[ \sum_i \bar p_i m_{i,1}\ell_i(t)\right] \\
&+2\mbox{Cov} \left[ \sum_i \varrho_i(t)m_i, \sum_i\bar p_i m_{i,1}\ell_i(t)\right]. 
\end{aligned}
\end{equation}

The correction terms \eqref{Bterm-1} and \eqref{Aterm-1} provide several insights into the separate and combined effects of variation in the frequencies of the dispersal models and the mean dispersal distances. Since the variance $\mbox{Var}\left[ \sum_i \varrho_i(t)m_i\right]$ in the frequencies of dispersal modes only show up in the second correction term \eqref{Aterm-1},  small fluctuations in these frequencies \emph{per-se} decrease invasion speeds. All else being equal, negative cross-correlations between dispersal modes wth similar kernels increase invasion speeds (i.e., if one mode of long-range dispersal is rare when another is common), while positive cross-correlations between similar modes decreases invasion speeds.  

The variances $\mbox{Var}\left[ \ell_i(t)\right]$ in mean dispersal distances appear in both correction terms \eqref{Bterm-1} and \eqref{Aterm-1}. The combined effect of these terms on the invasion speed equals
\be
\frac{\langle v, Fw \rangle}{s^*\lambda_0} \left( \sum_i \bar p_i m_{i,2} \mbox{Var}[\ell_i(t)]- \dfrac{\langle v, Fw \rangle}{2  \lambda_0}\mbox{Var} \left[ \sum_i \bar p_i m_{i,1}\ell_i(t)\right]\right). \label{eq:modes2}
\ee 
We show in  \ref{Appendix:mgf} that \eqref{eq:modes2} is always positive. Therefore, the net effect of the variability 
in mean dispersal distances is to increase the invasion speed. Moreover, equations \eqref{eq:modes3b} and \eqref{eq:modes2} implies that 
negative cross-correlations between the $\ell_i(t)$ result in a larger positive effect on the invasion speed $c^*$, 
while positive cross-correlations ameliorate this effect.

The net effect of fluctuations in the frequencies in the dispersal modes and their mean dispersal distances is determined by their separate effects plus the covariance terms in equations \eqref{Bterm-1} and \eqref{Aterm-1}. The combined effect of these covariance terms on the invasion speed equals 
\be
\begin{aligned}
& \frac{\langle v, Fw \rangle}{s^*\lambda_0} \left( \sum_i  m_{i,1}\mbox{Cov}[\varrho_i(t),\ell_i(t)]
- \dfrac{\langle v, Fw \rangle}{\lambda_0}\mbox{Cov} \left[ \sum_i \varrho_i(t)m_i, \sum_i\bar p_i m_{i,1}\ell_i(t)\right]\right)\\
%&= \frac{\langle v, Fw \rangle}{s^*\lambda_0} \left( \sum_i  m_{i,1}\mbox{Cov}[\varrho_i(t),\ell_i(t)]
%- \dfrac{\langle v, Fw \rangle}{\lambda_0}\bar p_i m_{i,1}\mbox{Cov} \left[ \sum_i \varrho_i(t)m_i, \ell_i(t)\right]\right)\\
&= \frac{\langle v, Fw \rangle}{s^*\lambda_0} \sum_i  m_{i,1}\mbox{Cov}\left[\varrho_i(t)
- \dfrac{\langle v, Fw \rangle}{\lambda_0}\bar p_i \sum_j \varrho_j(t)m_j, \ell_i(t)\right]
\end{aligned} \label{final-one}
\ee
While \eqref{final-one} shows that generally the correlations between frequency ($\rho_i(t)$) and range ($\ell_j(t)$) of dispersal 
potentially have complex interactive effects on the invasion speed, this interplay simplifies significantly when the $m_i$ are all equal e.g. all dispersal modes have the same expected mean dispersal distance and underlying movement kernel. In this case, the fact that $\sum_j p_j(t)=1$ implies that $\sum_j \varrho_j(t)=0$ and \eqref{final-one} simplifies to 
\be
\frac{\langle v, Fw \rangle}{s^*\lambda_0} \sum_i  m_{i,1}\mbox{Cov}\left[P_i(t), L_i(t)\right].
\label{final-one-really}
\ee
Thus, faster rates of spatial spread occur if a given dispersal mode has a higher mean dispersal distance when it is more common.

\begin{figure}[t]
\begin{center}
\includegraphics[width=0.95\textwidth]{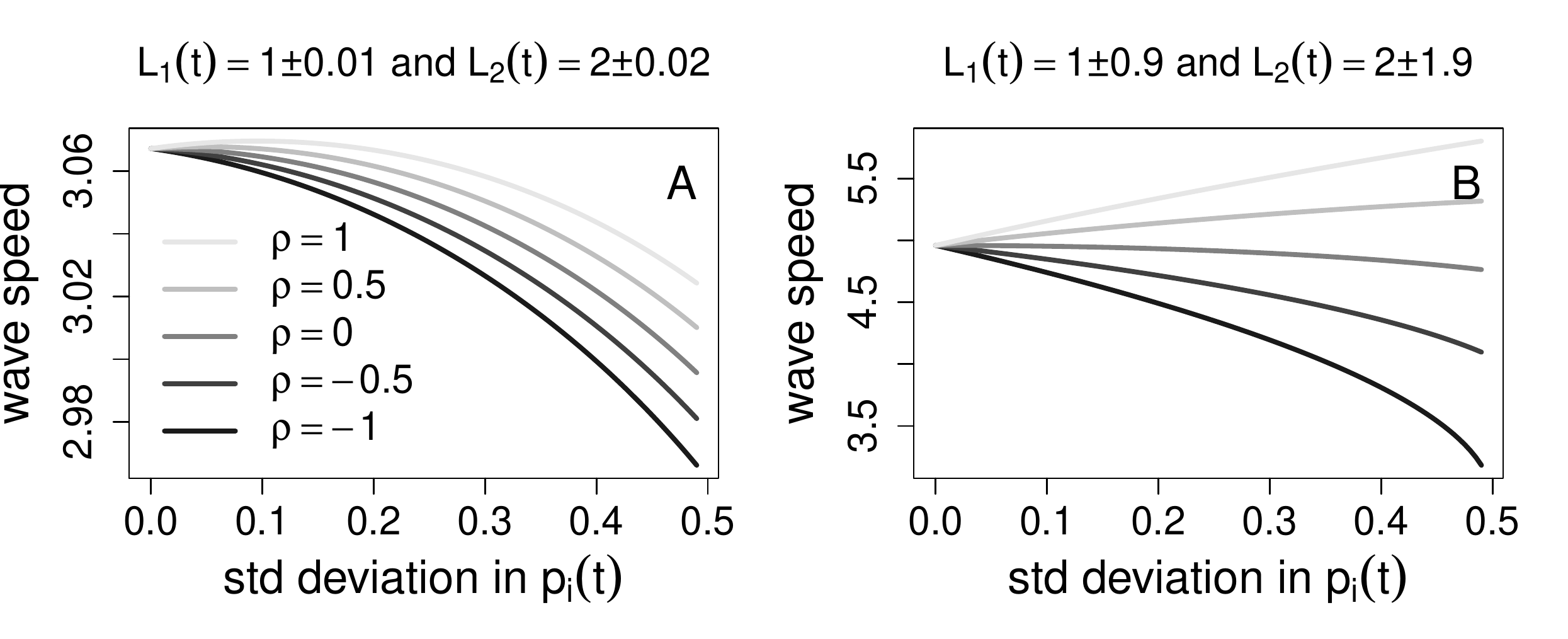}
\end{center}
\caption{Invasion speeds for an unstructured population with two modes of dispersal. Dispersal 
frequencies given by  $p(t)=p_1(t)$ which takes on the values $0.5\pm \sigma$ with equal probability, 
and $p_2(t)=1-p(t)$. The mean dispersal distances are given by $L_i(t)=\bar L_i \pm \tau_i$ with equal 
probability. The correlation between $p_i(t)$ and $L_i(t)$ is $\rho$, while the correlation 
between $p_i(t)$ and $L_j(t)$ with $j\neq i$ is $-\rho$. In  A, $\tau_1=0.01$ and $\tau_2=0.02$. 
In B, $\tau_1=0.9$ and $\tau_2=1.9$. In both figures, $P=0.6$, $F=1.5$, $\bar L_1=1$, $\bar L_2=2$, and $m_i(s)$ 
are Laplace moment generating functions. }
\label{fig:modes}
\end{figure}

To illustrate some of the implications of these approximations, we consider the simple case of an 
unstructured population exhibiting two modes of dispersal. The frequencies of these models are given 
by $p(t)=p_1(t)$ which takes on the values $0.5\pm \sigma$ with equal probability, and $p_2(t)=1-p(t)$. 
The mean dispersal distances are given by $L_i(t)=\bar L_i \pm \tau_i$ with equal probability. 
The correlation between $p_i(t)$ and $L_i(t)$ is $\rho$, while the correlation between $p_i(t)$ 
and $L_j(t)$ with $j\neq i$ is $-\rho$. Fig.~\ref{fig:modes}A illustrates, as asserted by 
approximation~\eqref{Aterm-1}, that larger fluctuations in the frequencies of the dispersal 
models decreases the invasion speed provided that fluctuations in the mean dispersal distances 
are sufficiently small. In contrast, when these fluctuations are sufficiently large and 
positively correlated with the corresponding dispersal mode frequencies, Fig~\ref{fig:modes}B 
illustrates,  larger 
fluctuations in the dispersal modes can increase invasion speeds. 

\subsection{Interactive effects of demographic and dispersal variation on spatial spread}
Temporally uncorrelated fluctuations in demographic rates (e.g. survivorship, fecundity), 
as discussed earlier,  reduce the stochastic growth rate $\log\lambda_S$ of a population and, 
as a consequence, reduce invasion speeds. However, when there are simultaneously fluctuations 
in demographic rates and mean dispersal distances, correlations between demography and dispersal
 may alter this conclusion. To investigate this possibility, consider the juvenile dispersal model with one mode 
 of dispersal and assume that the fluctuations in the mean dispersal distance $L_t$,  the fecundity 
 kernel $F_t$, and the survivorship and growth kernel $P_t$ are small. Define
\begin{eqnarray*}
m&=&M_1(\bar L s^*), \quad  m_1= M_1'(\bar L s^*)s^*,  \quad m_2 = M_1''(\bar L s^*)(s^*)^2/2,\\
\ell_t &=& L_t- \bar L, \quad  \pi_t = P_t -\bar P,   \quad \phi_t =F_t-\bar F,
\end{eqnarray*} 
where  $\bar L=\E[L_t]$ is the expected mean dispersal distance, $\bar F = \E[F_t]$ is the expected fecundity kernel,  $\bar P=E[P_t]$ is the expected survival/growth kernel, and $s^*$ determines the invasion speed $c^*=c^*(s^*)$ for the unperturbed kerned $H_s^0 = 
\bar P+ M_1(\bar L s) \bar F$. If $\ell_t$, $\pi_t$, and $\phi_t$ are small (i.e. of order $\varepsilon$), a Taylor approximation of the moment generating function up to order $\varepsilon^2$ yields
\begin{eqnarray*}
H_{s^*,t}&=& P_t + M_1(L_t s^*)F_t \\
&\approx & P_t+ \left(m+m_1 \ell_t +  m_2\ell_t^2 \right)F_t\\
&\approx &  \underbrace{\bar P + m \bar F}_{H_{s^*}^0} +\underbrace{ \pi_t+  m\phi_t+ m_1  \ell_t \bar F}_{A_t}
+\underbrace{m_1  \ell_t \phi_t+m_2\ell_t^2 \bar F}_{B_{t}}.
\end{eqnarray*}
The perturbation term $B_t$ has non-zero mean of order $\varepsilon^2$ and variance of order $\varepsilon^4$ and, consequently, this variance is negligible.  The perturbation term $A_t$ has mean zero and variance of order $\varepsilon^2$.

Let $\lambda_0$, $v$, and $w$ be the the dominant eigenvalue, the reproductive value and the stable state distribution for $H_{s^*}^0$, respectively. Using the sensitivity formula \eqref{eqn:s1} and the fact 
that $\langle v ,H^0_{s^*} w\rangle = \lambda_0$, the correction term to the invasion speed $c^*$ due to the perturbation $B_{t}$  is 
\begin{equation}
\frac{1}{\lambda_0 s^*}\E[\langle v, B_{t} w \rangle] = \frac{1}{\lambda_0s^*}\left( m_1 \mbox{Cov}[\ell_t , \langle v , \phi_t w \rangle] + m_2 \langle v ,\bar F w\rangle \mbox{Var}[\ell_t]\right).
\label{Bterm}
\end{equation}  
Since the perturbation $A_t$ has zero mean, the sensitivity formula \eqref{eq:s2} implies that the correction term to the invasion speed $c^*$ due to  the perturbation $A_{t}$  equals
\begin{equation}
\begin{aligned}
 - \dfrac{1}{2 s^* \lambda_0^2} \E [ \langle v, A_t w \rangle^2 ]
& = - \dfrac{1}{2 s^* \lambda_0^2} \mbox{Var} \left[  \langle v, \pi_t w\rangle + m \langle v ,\phi_t w\rangle + m_1 \ell_t \langle v, \bar F w\rangle\right]
\label{Aterm}
\end{aligned} 
\end{equation} 
where we can rewrite the variance term as 
\begin{equation}
\begin{aligned}
\mbox{Var} \left[  \langle v, \pi_t w\rangle + m \langle v ,\phi_t w\rangle + m_1 \ell_t \langle v, \bar F w\rangle\right]=&
\mbox{Var} \left[  \langle v, \pi_t w\rangle + m \langle v ,\phi_t w\rangle \right] + m_1^2\langle v, \bar F w\rangle^2\mbox{Var}\left[  \ell_t  \right]\\
&+2m_1\langle v, \bar F w\rangle\left(  \mbox{Cov}\left[ \langle v, \pi_t w\rangle, \ell_t  \right] +m\, \mbox{Cov}\left[  \langle v ,\phi_t w\rangle, \ell_t  \right] \right).
\end{aligned}
\label{Aterm-var}
\end{equation}

The correction terms \eqref{Bterm} and \eqref{Aterm} provide several insights into the separate and joint effects of variation in dispersal and demography on invasion speeds. The variance in the mean dispersal distances $\mbox{Var}[\ell_t]$ occurs in both correction terms \eqref{Bterm} and \eqref{Aterm}. However,  
the fact that $M_1''M_1 > (M_1')^2$ implies, just as argued in Section 5.2, that the net effect of dispersal variation \emph{per se} 
is to increase invasion speed.  Demographic variance occurs only in the second correction term \eqref{Aterm}. Consequently, the effect of variance in survival or 
fecundity \emph{per se} is to decrease the invasion speed by 
\begin{equation}
-\frac{1}{2\lambda_0^2 s^*} \mbox{Var} \left[  \langle v, \pi_t w\rangle + m \langle v ,\phi_t w\rangle \right].
\label{dd1}
\end{equation}
 This effect is largest if survival and fecundity are positively correlated, 
and smaller if their correlation is negative.    

The covariance $\mbox{Cov}[ \langle v , \pi_t w \rangle,\ell_t ]$ between survival and mean dispersal distance occurs only in the second correction term \eqref{Aterm}. Consequently, all else being equal, a positive dispersal-survival covariance decreases the invasion speed, while 
a negative correlation increases the invasion speed. The covariance $\mbox{Cov}[\langle v, \phi_t w \rangle, \ell_t]$ between fecundity and mean dispersal distance  occurs in both correction terms \eqref{Bterm} and \eqref{Aterm}. The net effect of this covariance from both correction terms is 
\begin{equation}
\frac{m_1}{s^* \lambda_0}\mbox{Cov}[\langle v, \phi_t w \rangle, \ell_t]\left(1-\frac{\langle v, m \bar F w\rangle}{\lambda_0}\right)
=\frac{m_1}{s^* \lambda_0} \mbox{Cov}[\langle v, \phi_t w \rangle, \ell_t] \underbrace{\langle v, \bar Pw \rangle}_{\ge 0}. 
\label{DFcorr}
\end{equation} 
Demographic variability therefore makes two contributions: one negative (equation \eqref{dd1}) and
the other positive when fluctuations in fecundity and dispersal are positively correlated 
(equation \eqref{DFcorr}). If the second of these is the larger in magnitude, the net effect of demographic variability
is to increase invasion speed, a phenomena demonstrated numerically in our analysis  of a 
patch-based model of pepperweed spread (see section 7 and Fig~\ref{fig:Pepperweed1}). Equation~\eqref{DFcorr} also implies that  for  annuals (i.e. $P=0$), the covariance between dispersal and fecundity has no effect on the invasion speed. In fact this observation holds in general in our plant model with juvenile-only dispersal, not just for small fluctuations. 
With $H_{s,t}=M_1(L_t s)F_t$, the Lyapunov exponent $\gamma(s)$ is the sum of 
$\E \log M_1(L_t s)$ and the dominant Lyapunov exponent for $\{F_t\}$. Hence, covariance between $L_t$ and $F_t$ 
has no effect on the invasion speed for annuals. 

\section{Is dispersal variability important? }
\label{sect:DoesItMatter}
\cite{andrew-ustin-10} used remote-sensing data to estimate four interannual dispersal kernels 
for patches of perennial pepperweed, \textit{Lepidium latifolium}, in the Sacramento-San 
Joaquin River delta in California, and developed a simulation model for pepperweed spread. 
Dispersal differed among habitats and was also highly variable from year to year. While most new recruits 
established close to their parents in all years, long-range dispersal ($>100$~m) occurred in 
two interannual periods ($\approx$ 0.2\% and 2\% of new recruits). As a result, the average dispersal
distance (combining all sites and habitat types for each year) had a temporal coefficient of variation of 
approximately 40\% \citep[][Table 3]{andrew-ustin-10}. 

However, they noted that ``invasion dynamics are little affected by temporal variation in
dispersal distances. This contradicts theoretical expectations
from analytical models, which predict that temporal variation in either reproductive rate or
dispersal will reduce spread rates to the geometric mean
of the rates observed under the component constant conditions (Neubert et al. 2000).''
Specifically, simulated rates of spread with temporally variable dispersal were very
similar to the rate when the dispersal kernel with the farthest long-distance
dispersal was used in all years (their Figure 6). This appears to to contradict our analytic
results above on the importance of dispersal variation, as well as previous theory. 

In this section, we explain why the findings of \citet{andrew-ustin-10} are consistent
with our results and with previous theory, and actually are predicted by our results. 
The explanation is a phenomenon that has been observed 
numerically in models with temporally constant dispersal:
rates of spread are far more sensitive
to the \textit{range} of long-distance dispersal than to the \textit{fraction} of long-distance dispersers
(see Figs. 13 and 14 in \citet{neubert-caswell-00}). The corresponding 
phenomenon for temporally variable dispersal is that the rates of spread are insensitive
to the fraction of years in which long-distance dispersal occurs, and are largely determined
by the mean dispersal distance in those years. 

\begin{figure}[tbp]
\centering
\includegraphics[width=0.9\textwidth]{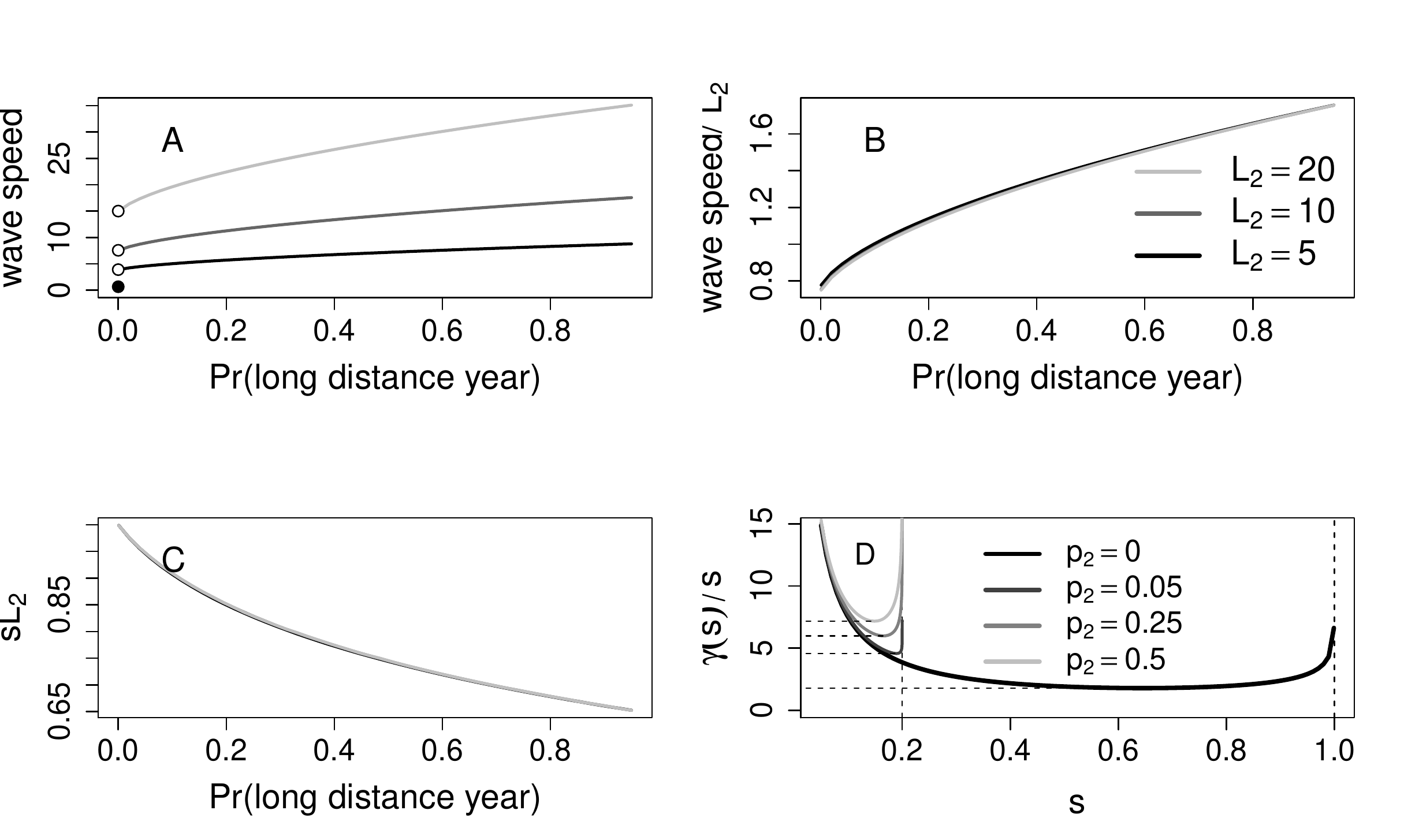}
\caption{Numerical calculations of wave speed for the unstructured perennial plant population described in the text,
with parameter values $L_1=1, F=1.5, P=0.6.$ (A) Asymptotic wave speed $c^*$ as a function of $p_2$, the probability
of a long-distance dispersal year, for $L_2=5,10,20$ (see legend of panel B). 
The curves show $c^*$ for $0.001 \le p_2 \le 0.95$; the solid 
circle is $c^*$ for $p_2=0$, which does not depend on the value of $L_2$. (B) $c^*/L_2$, the asymptotic wave speeds relative
to the scale of long-distance dispersal. (C) $s^* L_2$, the asymptotic wave-shape parameters corresponding 
to $c^*$. (D) The graph of $\gamma(s)/s$ for $L_2=5$ and various values of $p_2$. 
The minimum values of $\gamma(s)/s$, indicated by horizontal dashed lines, correspond to the asymptotic 
wave speed $c^*$. The vertical dashed lines correspond to the positive vertical 
asymptotes of $\gamma(s)/s$ when $p_2=0$ and when $p_2>0$.  }
\label{fig:UnstructuredWaveSpeed}
\end{figure}

In Figure \ref{fig:UnstructuredWaveSpeed}, we illustrate this phenomena for an unstructured population with juvenile dispersal. The transformed kernel 
is $M_t(s)F+P$ where $F$ is the per-capita fecundity, $P$ is annual survival of adults, and $M_t(s)$ is the moment
generating function for the dispersal kernel in year $t$. The dispersal kernel is a Laplace (two-sided exponential)
kernel with two possible distance parameters $L_2 > L_1$ representing long- and short-range dispersal years.  
The temporal variability was uncorrelated, the distance parameter in each year being $L_1$ with probability
$p_1$ and $L_2$ with probability $p_2=1-p_1$. Figure \ref{fig:UnstructuredWaveSpeed}A shows that the invasion speed $c^*$ changes discontinuously
at $p_2=0$: a minuscule chance of long-range dispersal (in these computations, $p_2=0.001$) is very different
from no chance at all. But a hundred-fold further increase in the chance of long-range dispersal (from $0.001$
to $0.1$) has relatively little effect, especially when $L_2 \gg L_1$. Figures \ref{fig:UnstructuredWaveSpeed}B and \ref{fig:UnstructuredWaveSpeed}C show that to a very good approximation, the wave speed $c^*$ and the corresponding value of $s^*$ that minimizes $\gamma(s)/s$ are  linearly proportional to $L_2$ (so long as $p_2>0$), but are only weakly dependent on  $p_2$. 

The geometric underpinnings of this discontinuity, as illustrated in \ref{fig:UnstructuredWaveSpeed}D, are threefold. 
First, $M_t(s)$ is defined on the interval $(0,1)$ with probability one for $p_2=0$. However, when there is a 
positive probability $p_2>0$ of long distance dispersal with $L_2=5$, $M_t(s)$ is defined with probability one only 
on the smaller interval $(0,0.2)$. Second, when $p_2=0$, 
the minimal value of $\gamma(s)/s$ occurs at $s^*\approx 0.6 > 0.2$, while with $p_2>0$ the minimizer of $\gamma(s)/s$ 
must occur in $(0,0.2)$. Finally, $\gamma(s)$ for $p_2>0$ is greater than $\gamma(s)$ for $p_2=0$. 
Collectively, these properties imply a discontinuous increase in the invasion speed at $p_2=0$.

Although we have used an unstructured model for simplicity and illustrative purposes, this phenomenon is quite general. For example, suppose that $H^i_{s,t}(z,z_0)$ with $i=1,2$ are general transformed kernels of the form \eqref{eqn:skernels} such that $H^i_{s,t}$ are well-defined with probability one on the maximal interval $(-\hat s_i,\hat s_i)$ with $\hat s_1$ possibly infinite. Let $H_{s,t}$ equal $H^1_{s,t}$ with probability $p_1$ and $H^2_{s,t}$ with the complementary probability $p_2=1-p_1$. If $\gamma_1(s)/s$ is minimized at $s_1^*>\hat s_2$ and $\gamma_2(s) \ge \gamma_1(s)$ for $0<s<\hat s_2$, then there is a jump discontinuity in the invasion speed at $p_2=0$. Moreover, a lower bound for the size of this jump discontinuity is $\gamma_1(\hat s_2)/\hat s_2 - \gamma_1(s_1^*)/s_1^*$. Since this argument relies on $\hat s_2$ being finite, these jump discontinuities do not occur for thin-tailed dispersal kernels such as a Gaussian dispersal kernel. 

\section{Application: a patch-based model for pepperweed spread}
To illustrate our results and the role of variable dispersal we develop 
and study a simple model for the spread of perennial pepperweed. Pepperweed is a Eurasian crucifer that has
spread widely in the western United States since it was introduced into the US in the early
20th Century. It is now invasive in wetlands and riparian zones throughout the western 
US \citep{leininger-foin-09}, and an agricultural weed of concern to 
alfalfa and native hay growers \citep{blank-etal-02}.  
The model is based on population studies in the Cosumnes River Preserve 
(Hutchinson et al. 2007, Viers et al. 2008) and nearby sites in the Sacramento-San 
Joaquin River delta, California \citep{andrew-ustin-10}. 
Pepperweed has been identified as a significant threat in the Preserve because it is highly invasive there and forms 
monospecific stands where native species are completely eliminated.  
Our model incorporates the main features of pepperweed's life history and 
dispersal, but given the limited information on some demographic processes we can only claim that the model is
inspired by pepperweed. The main purpose of the model is to explore interactions between temporal
variation in local demography and dispersal. 

\subsection{Pepperweed life history} 
Rosettes develop in the spring and eventually bolt to produce multiple inflorescences 0.5-2m tall 
\citep{leininger-foin-09}. Seed production can be quite high, with 
high germination and viability (roughly 3000 seeds/inflorescence and
174,000 seeds/m$^2$ within a patch in the Preserve, with 
96.4\% germination and over 80\% viability seven months after seed production
\citep{leininger-foin-09}). However seed production and viability were 
substantially lower in moister or more saline sites \citep{leininger-foin-09}. 

The root system of established plants spreads laterally and produces new sprouts, 
creating a dense local patch from which all other species are eventually excluded. 
In suitable habitat, rhizomes can spread 1-2 meters within a year \citep{orth-etal-06},
and a single plant can grow into a patch several meters
in diameter within 2 years \citep{leininger-foin-09}. Root fragments can also 
disperse and resprout to create new patches. Patches are difficult to 
eradicate mechanically because of the deep root system and resprouting of new 
plants from root fragments. Pepperweed also forms a long-term seed bank, which can 
re-establish a patch after all adult plants have been killed \citep{viers-etal-08}. 

Water -- soil moisture and flooding -- has been identified as the main environmental
driver of demographic variability. The study sites are seasonally flooded each year. 
In the Cosumnes River Preserve, established patches tend to shrink during wet 
years (i.e., growing seasons following a wet spring), and grow during 
dry years \citep{hutchinson-etal-07}. Patches can disappear in a wet year, but 
reappear in a subsequent dry year.
At the ``Experimental Floodplain'' site in the Preserve, 153 patches (out of 443 total) 
disappeared between 2004 and 2005 and 40 (out of 312 total) disappeared between
2005 and 2006, but 75\% of those had reappeared by 2007 \citep{hutchinson-etal-07}.
New patches appeared mainly during dry years, but it is
believed that these were initiated by seeds or root fragments that had been
moved by water flow in wet years \citep{hutchinson-etal-07}. However, the effect
of rainfall and flooding is site-dependent. During the same time period, at Bouldin Island 
nearby in the Sacramento-San Joaquin river delta, establishment of new patches 
was greatest in wetter years, with the largest effect in sites expected to be most 
water-limited \citep{andrew-ustin-10}. 

\subsection{Model}
Because pepperweed occurs mainly in dense monospecific stands, we model  
the growth and spread of patches rather than individual plants. This allows
us to use a density-independent model for the growth and spread of patches. Even though
most plants experience strong intraspecific competition from others in their
patch (even at the leading edge of the invasion), a density-independent model 
of patch dynamics can be constructed for the situation where patches are sparse and inter-patch
competition is rare. This model becomes inaccurate once 
patches are so big and common that they start to collide. But according to the linearization
hypothesis, the linear model that applies to sparse patches at the leading edge of the
invasion correctly predicts the asymptotic spread rate. 

We assume the basic ``plant'' model, equation \eqref{eqn:PlantKernel}, approximating patches 
as expanding circles. Established patches have been reported to increase in diameter 
by 1-3 m/yr (Andrew and Ustin 2010), which is comparable to the initial rate of 
expansion from a founding plant. We therefore 
assume that patches grow at mean rate $g(t)$ meters/yr, where $g(t)$ 
depends on environmental conditions in year $t$, but not on patch size. To allow variance
in growth among patches, we assume a lognormal distribution of growth rates with variance
$\sigma^2_{g}$, assumed to be constant. The log-transformed radius of new patches
was assumed to follow a beta(3,3) distribution shifted to have mean -3; because of the
rapid patch growth, the initial size distribution has essentially no effect on model predictions. 
Patches ``die'' by disappearing in a wet year and never reappearing, and we let $d(t)$ denote the 
annual probability of patch ``death'' (in the Experimental Floodplain data, $\E[d] \approx 0.06$). 

We assume a Laplace (bi-exponential) dispersal kernel with time-varying mean
dispersal distance $L(t)$. This was based on the summary data in Table 3 of Andrew and Ustin (2010), 
which showed that for each habitat-year combination the mean and standard deviation of dispersal 
distance are roughly equal, and the observed maximum distance is roughly what would be expected
under an exponential distribution given the mean distance and the sample size. We assume that $L(t)$ has a mean of 15m 
(roughly the average over all years and habitats in Table 3 of \cite{andrew-ustin-10}). We
model one-dimensional spread, e.g., expansion along a strip of riparian habitat.  

We assume that fecundity -- the founding of new patches by propagules from an established patch -- is 
proportional to patch radius squared (i.e. to patch area), with time-varying constant 
of proportionality $f(t)$.  

Let $n(\rho,x,t)$ be the density at location $x$ in year $t$ 
of pepperweed patches of size $\rho$, where $\rho$ is the natural log of plant radius. The
demographic kernels making up the model are then 
\begin{equation}
\begin{aligned}
k_{d,t}(v) & =\frac{1}{2L(t)}e^{-|v|/L(t)} \\
F_t(\rho,\rho_0) & = f(t)e^{2 \rho_0}\beta_{3,3}(3+\rho) \\
P_t(\rho,\rho_0) & = (1-d(t))\phi(\rho;\log(e^{\rho_0}+g(t)),\sigma^2_g) 
\end{aligned}
\label{eqn:Pweedkernel}
\end{equation}
where $\phi(\bullet;\mu,\sigma^2)$ is the Gaussian density with mean $\mu$ and variance $\sigma^2$. 
The nonspatial kernel describing total population growth is then $K_t=F_t + P_t$, and the transformed kernels
that determine the asymptotic spread rate is $H_{s,t}=(1-s^2L(t)^2)^{-1}F_t + P_t.$  
Because the model \eqref{eqn:Pweedkernel} is formulated in terms of continuous growth kernels, 
the demography is described by just a few parameters, which is very helpful for analyzing 
the model's behavior. 

\subsection{Pepperweed model results}

\begin{figure}[tbp]
\centering 
\includegraphics[width=0.9\textwidth]{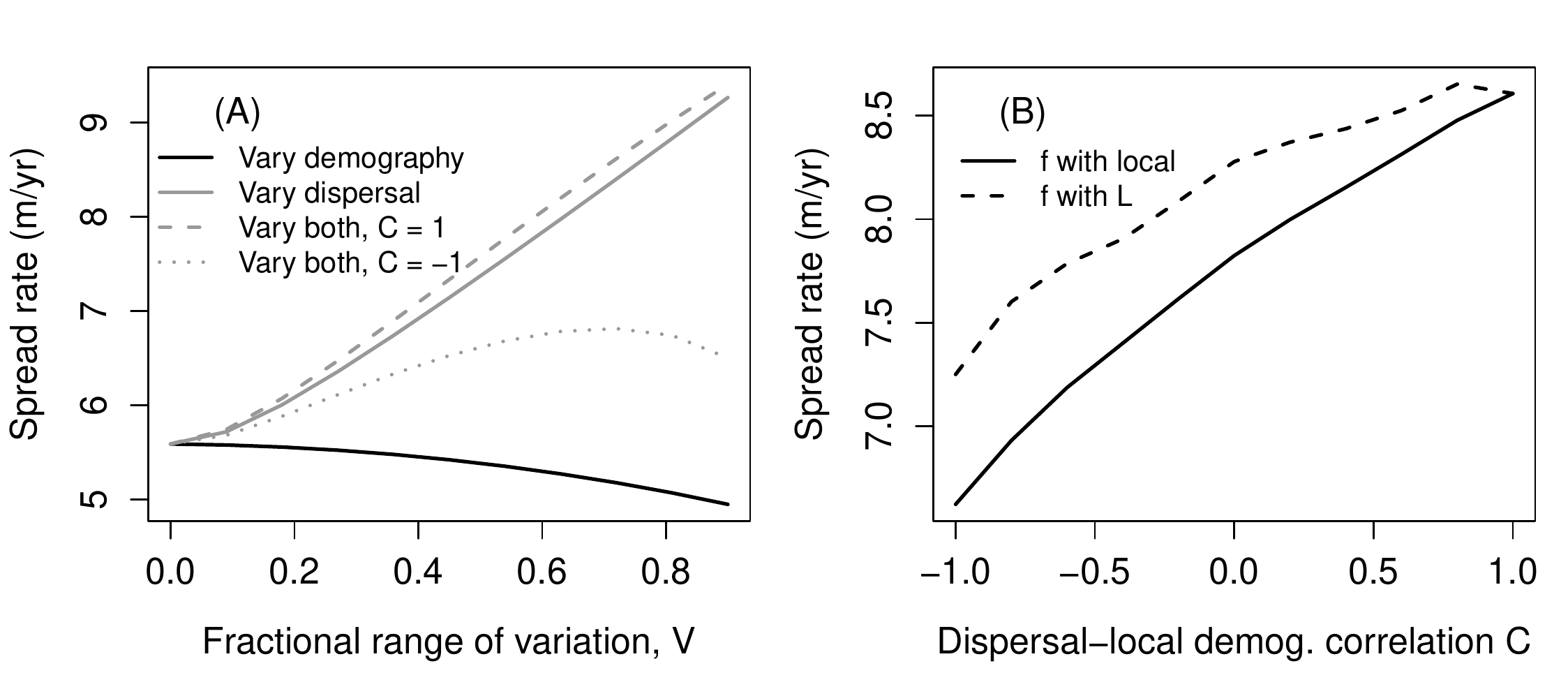}
\caption{Numerical calculations of wave speed for the perennial pepperweed model. 
Mean values of the model parameters were 
$\bar L=15, \bar g=0.5, \bar d =0.06, \bar f =0.04$ ($\sigma_g$ was held constant).
(A) Temporal variability consisted of equally likely wet and dry years characterized
by the fractional range of variability $V$, meaning that the two possible values
of a parameter $\theta$ are $\bar \theta \pm V \bar \theta$ ($\theta=L,\gamma, d,$
and/or $f$). ``Vary demography'' (solid black curve): $L$ is constant, all other
parameters vary in parallel: rapid growth, high fecundity, and low mortality during
wet years. ``Vary dispersal'' (solid gray curve): $L$ varies (further dispersal in
wet years), all other parameters are constant. ``Vary both'': all parameters vary; with
correlation $C=1$, wet years are good for both local demography and long-distance dispersal,
and with correlation $C=-1$ wet years are good for long-distance dispersal but bad for
local demography. (B) Temporal variability with four year types, characterized by all possible
combinations of high or low long-range dispersal ($L=\bar L \pm V \bar L$) and good or poor local
demography (e.g., $g = \bar g \pm V \bar g$) with $V=0.8$. $C$ is the degree of 
correlation between fluctuations in local demography and dispersal. Solid curve: fecundity parameter
$f$ is considered to be a part of local demography. Dashed curve: $f$ is associated with dispersal,
so $f$ and $L$ vary in parallel.}
\label{fig:Pepperweed1}
\end{figure}

We adjusted the mean of the fecundity coefficient $f(t)$ to match the
observed average ``birth rate'' between 2002 and 2007 in the Experimental Floodplain site
(0.3 new patches per established patch per year). This indirect estimate is supported
by the fact that the model then predicts a long-term population growth rates of $\lambda_S = 1.20 - 1.24$
depending on the level of stochasticity, which matches very well the the observed rate of increase
in the total number of patches in the Experimental Floodplain site ($(456/158)^{1/5} \approx 1.24$). 

Figure \ref{fig:Pepperweed1}A illustrates the effect of dispersal variability alone and in combination
with demographic variability. These calculations used a simple wet years/dry years pattern
of environmental variation, with the fractional level of variation $V$ running from 0 to 90\% of
the maximum possible (e.g., $V=0.5$ means that the two (equally likely) possible values of the fecundity 
parameter $f$ are $\bar f \pm 0.5\bar f$, and similarly for the patch mortality $d$, patch
growth rate $\gamma$ and dispersal range parameter $L$). Variation in dispersal alone (only $L$ varies)
increases the asymptotic spread rate, as predicted by our perturbation analysis, while variation in
local demography alone decreases the spread rate ($L$ constant, all other parameter varying).
The small fluctuations approximation (equation \ref{eqn:SmallVar}) implies that small temporally uncorrelated 
local demographic variation always decreases the spread rate, regardless of how demographic parameters
covary. For our pepperweed model this remains true even when the difference between 
wet and dry years is very large. 

More interesting is the interaction between the two types of variation. If local demographic
variation is positively correlated with dispersal variation, so that wet years
are good for both local patch growth and long-distance dispersal, then 
the effect of demographic variation is reversed, as predicted by equation \eqref{DFcorr}: demographic
variation now increases the spread rate $c^*$, rather than decreasing it. 
Conversely, if the correlation between local demography and dispersal is negative,
local demographic variability decreases $c^*$. In Fig. \ref{fig:Pepperweed1}~A, patch fecundity
was regarded as a component of local demography, so when $C=-1$ the years when propagules
go far also tend to be years with fewer propagules. If we instead regard fecundity as a component
of dispersal (so that $L$ and $f$ vary in parallel), adding local demographic variation on top of
dispersal-distance variation still can increase the spread rate, but the effect of correlation
is reversed: spread rate is increased by the addition of demographic variation 
when $C = -1$, and decreased when $C=1$. Positive correlation between variation in local
demography and dispersal generally increases the spread rate (Figure \ref{fig:Pepperweed1}~B), 
regardless of whether $f$ is associated with local demography (i.e., high in years when survival 
and growth are high) or with dispersal (i.e., high in years when $L$ is high). Spread rate
is higher when fecundity and dispersal distance vary in parallel. This conforms to the
general principle that spread rate is largely determined by the very best years for long-range
dispersal, so making those years even better increases the spread rate even
though overall fecundity is unchanged.

\begin{figure}[tbp]

\centering 
\includegraphics[width=0.7\textwidth]{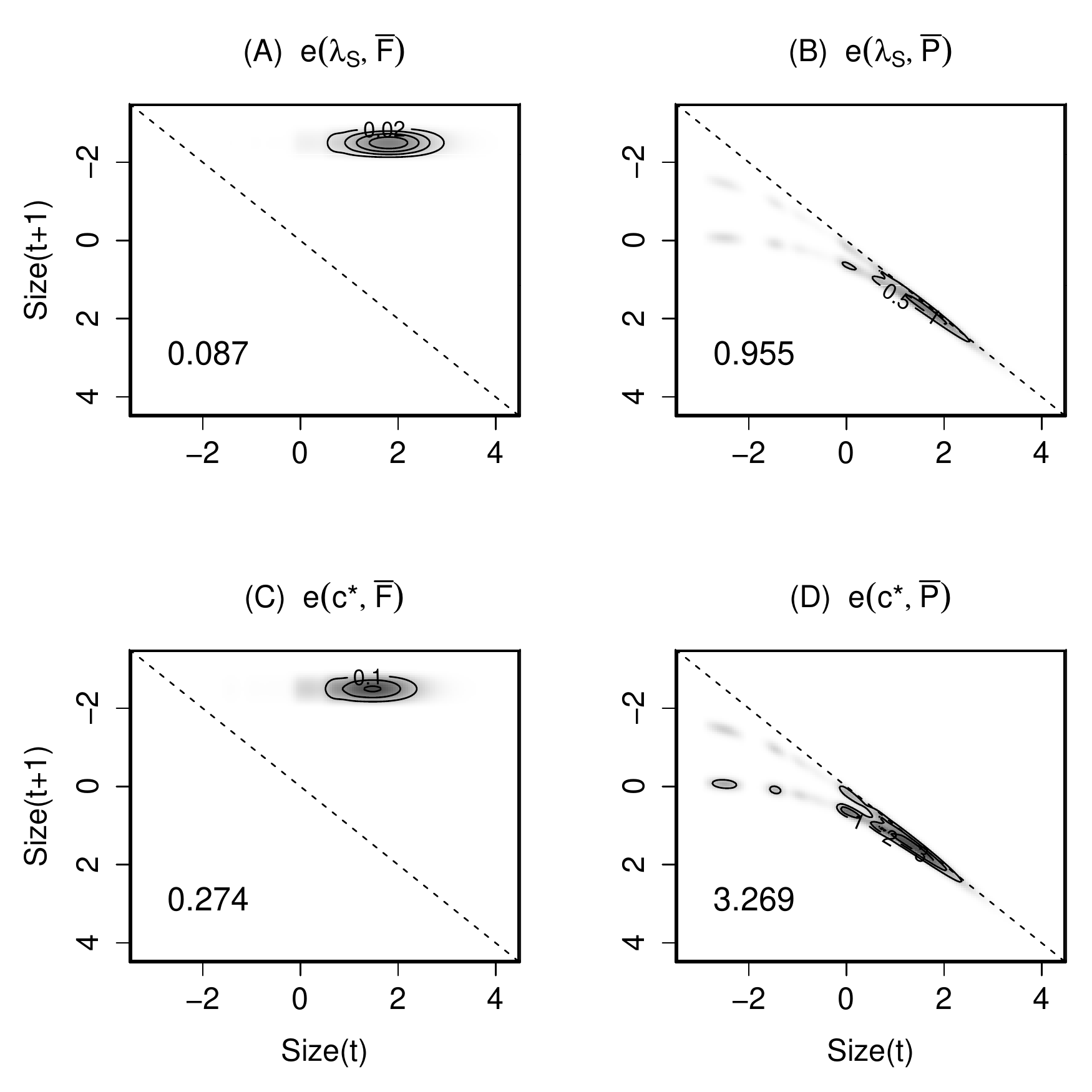}
\caption{Elasticities with respect to mean kernel entries for the perennial pepperweed model
with wet and dry year-types. The parameter values were the same as in Figure \ref{fig:Pepperweed1},
with fractional range of variation $V=0.7$, and correlation $C=-1$ so that wet years are good
for long-range dispersal but bad for local demography (including fecundity), relative to dry years. The four
panels show the elasticity surfaces $e(y,x)$ for the stochastic population growth rate 
$\lambda_S$ and the wave speed $c^*$ with respect to the mean values of $F(y,x)$ 
and $P(y,x)$. Elasticity values are indicated by the contour lines, and by shading from
zero (white) to high (dark grey). Numebers in the bottom-left corner are the total elasticity, i.e., the integral
of the plotted surface.} \label{fig:PepperweedElas1}
\end{figure}

\begin{figure}[tbp]
\centering 
\includegraphics[width=0.7\textwidth]{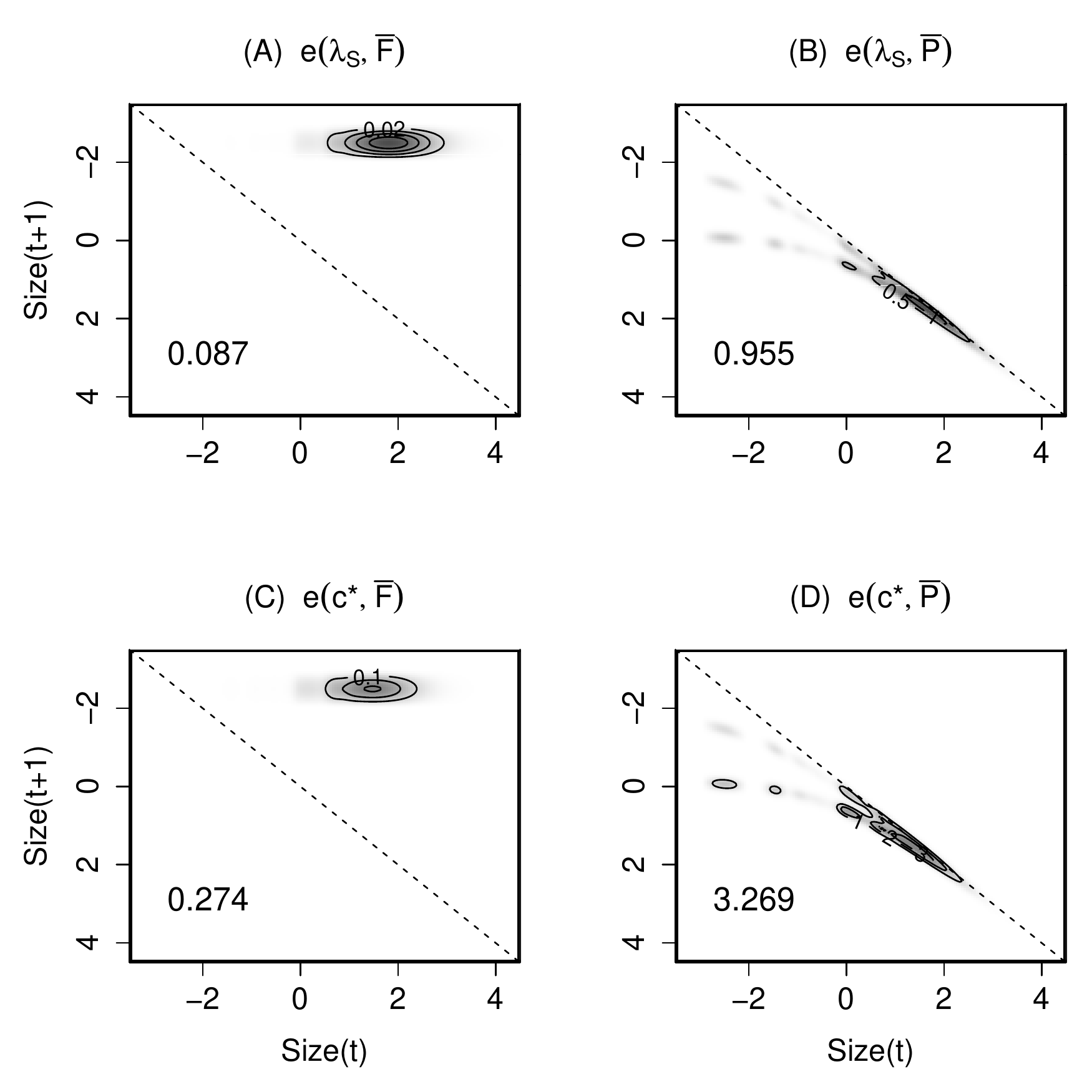}
\caption{The same as in Figure \ref{fig:PepperweedElas1} but with correlation $C=1$ so
that wet years are good for both long-distance dispersal and local demography, relative
to dry years.} \label{fig:PepperweedElas2}
\end{figure}

\begin{figure}[tbp]
\centering 
\includegraphics[width=0.5\textwidth]{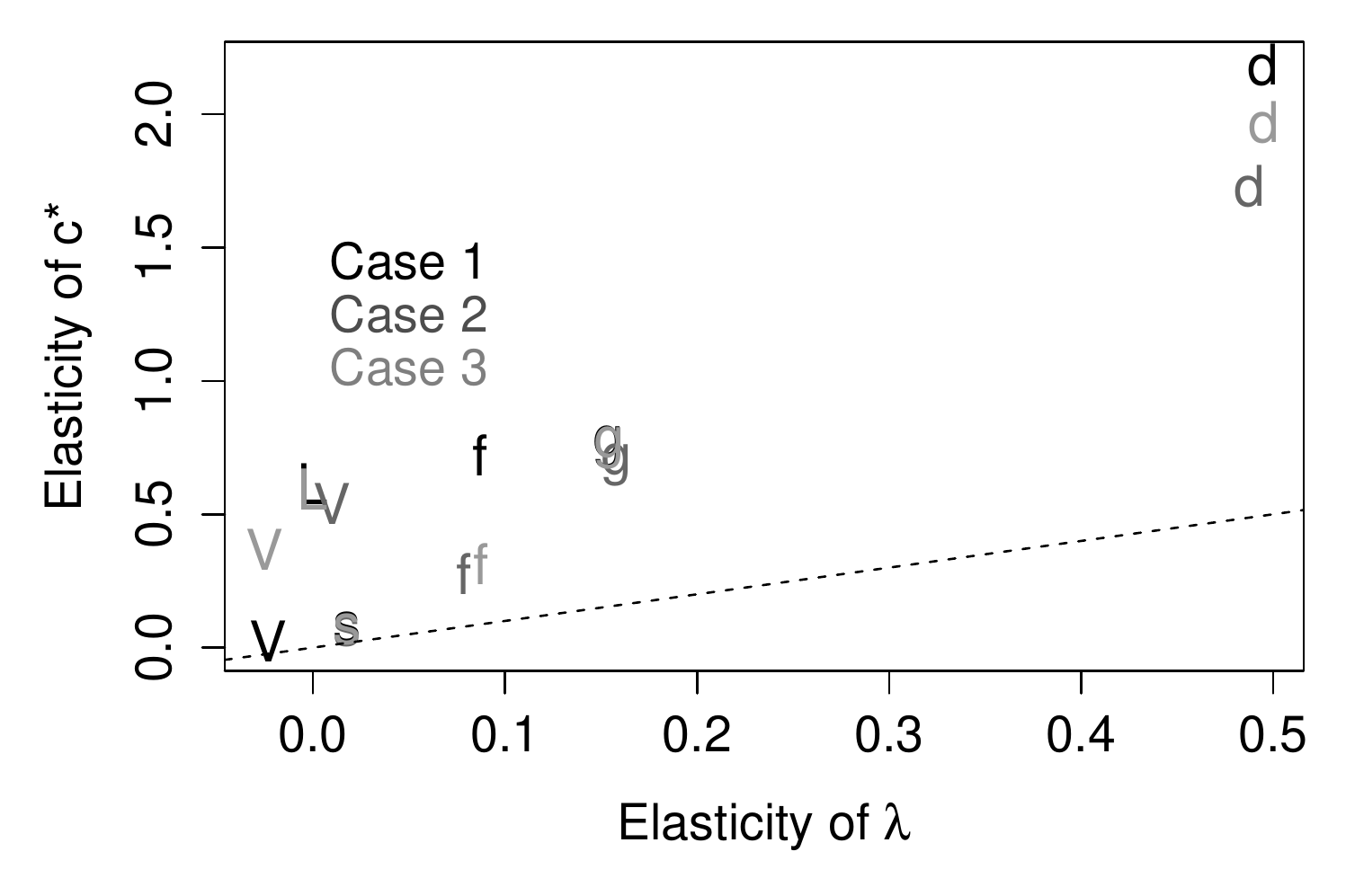}
\caption{Elasticities of population growth rate $\lambda_S$ and spread rate $c^*$ 
with respect to the mean of underlying model parameters $f$, $g$ and 
$\sigma_g$ (indicated by the symbol ``s'' in the plot). The parameter values were the 
same as in Figure \ref{fig:Pepperweed1}, with two year types and 
fractional range of variation $V=0.7$. In Case 1 (black symbols) wet years were assumed to be good for
long-range dispersal and poor for local demography including fecundity $f(t)$. In Case 2 
(darker gray symbols) wet years were assumed to be good for long range dispersal and fecundity,
and bad for patch growth and survival. In Case 3, wet years were assumed to be good for all
demographic processes. The dashed line is the 1:1 line.} \label{fig:PepperweedParamElas1}
\end{figure}

Figure \ref{fig:PepperweedElas1} show contour plots of the elasticity surfaces
for population growth rate $\lambda_S$ and wave speed $c^*$ in response to changes in
the mean value of a size-specific fecundity $F$ or survival/growth $P$, for a
wet/dry years scenario with negative correlation between local demography and dispersal 
(wet years are good for long-range dispersal, but bad for survival and fecundity). For example, 
the value of $e(c^*,F)$ at sizes $\rho_t=\rho_1,\rho_{t+1}=\rho_2$ is
the fractional change in $c^*$ per fractional change in the rate at which 
size $\rho_1$ parents produce offspring in the size range  $[\rho_2, \rho_2 + d\rho]$.
Numbers in the bottom-left corner of each panel
are the total elasticity (the integral of the plotted surface). For both $\lambda_S$ and $c^*$, 
the survival elasticities are higher, a typical pattern in long-lived organisms because 
an increase in adult survival leads to a large increase in lifetime reproductive success.   
Total elasticities are higher for $c^*$ than for $\lambda_S$. The total elasticity of $\lambda_S$
is necessarily near 1 (in a deterministic model the sum of the $P$ and $F$ total elasticities 
for $\lambda$ is exactly 1, here it is slightly different because of stochasticity). The higher 
total elasticity for $c^*$ indicates that a small change in the total population growth
rate produces a larger change in the spread rate. 

The relative unimportance of fecundity is less severe for $c^*$ than for $\lambda_S$ (panels
C and D). This is again not unexpected -- because it is only new offspring that disperse in
this model -- but it depends on the correlation between dispersal and local demographic
variation. When we assume instead that wet years are good for both local demography and
long-range dispersal, the relative importance of fecundity and survival/growth is about
the same for $c^*$ as it is for $\lambda_S$ (Figure \ref{fig:PepperweedElas2}).

We also computed the elasticities of $\lambda_S$ and $c^*$ to 
perturbations of the mean values of the underlying model parameters $f,g,d,\sigma_g$
and the range of variation $V$ (Figure \ref{fig:PepperweedParamElas1}). 
Again, $c^*$ is more sensitive than $\lambda_S$ to model perturbations, 
and the most important parameter by far is the mortality rate of established
patches, $d$ (the graph shows the elasticity to a proportional decrease in $d$, so
the effect is positive). The parameter elasticities are mostly 
insensitive to the pattern of correlation between local demography and dispersal, with two 
exceptions: the elasticity to $f$ was higher, and the elasticity to $V$ was lower,
in the case that wet years are good for dispersal (high $L$) but bad for growth, 
survival, and fecundity. Increased variability $V$ always increases spread rate,
but has very little effect on population growth rate. 
The positive effect of variance in patch growth reflects the nonlinear
relationship between patch size and fecundity, so that the gain from above-average patch 
growth more than offsets the loss from below-average growth. 

\section{Discussion}
Stochastic fluctuations in demographic rates, such as survivorship, growth, reproduction, and dispersal are pervasive
in natural populations and can have profound impacts on local population growth and persistence~\citep{tuljapurkar-90,boyce-etal-06,ellner-rees-07,tpb-09}. Recently the implications of these fluctuations for rates of range expansion have been examined for discretely structured  populations~\citep{caswell-etal-11,schreiber-ryan-11}. Here we have extended these studies, by allowing continuous structure (e.g., size) within populations, and by analyzing how interannual fluctuations in dispersal distance can affect the rate of spatial spread. Our paper continues the ongoing effort of extending the seminal work of \cite{kot-etal-96}, who introduced to ecology the use of discrete-time integrodifference models for population spread with general (non-Gaussian) dispersal kernels. Using the formalism of demographic integral projection models~\citep{easterling-etal-00,ellner-rees-06}, we combined general models of local demography incorporating continuous and discrete population structure with general dispersal distributions (potentially depending on the state of the dispersing individual and its parent), and general patterns of stationary environmental variation. 

For these general models, we show that the size of a species range asymptotically increases linearly with time. We provide an explicit expression for this invasion speed and, thereby, extend earlier results for unstructured populations~\citep{neubert-etal-00} and discretely structured populations~\citep{caswell-etal-11,schreiber-ryan-11}. We also derive sensitivity formulas for these invasion speeds to small perturbations in demography and dispersal and, thereby, extend work of \citet{caswell-etal-11} to populations with continuous as well as discrete structure. From the analytic perspective, these sensitivity formulas allow us to explicitly investigate how small temporal variation in dispersal distances and local demography influence invasion speeds. From an applied perspective, these sensitivity formulas provide methods to evaluate how uncertainty in parameter estimates yield uncertainty in estimates of invasion speeds, and to identify which parameters have the largest impact on invasion speeds. 

Using the sensitivity formulas, we show that stochastic fluctuations in mean dispersal distances increases invasion speeds. In sharp contrast, temporally uncorrelated fluctuations in vital rates associated with always local demography decrease invasion speeds. This latter effect follows almost directly from classical results in stochastic demography~\citep{lewontin-cohen-69,tuljapurkar-90,ellner-rees-07} where the population long-term growth rate is reduced by one-half of the ``net'' demographic variance experienced by the population (see e.g. equation (8) in \citet{rees-ellner-09}, which is the IPM version of equation equation (14.68) in \citet{caswell-2001}, Tuljapurkar's small variance approximation for temporally uncorrelated environments). Roughly, this reduction stems from population growth rate being a concave function of local demographic rates. Hence, by Jensen's inequality variation in these rates decreases the population growth rate and, hence, the invasion speed. In contrast, for the models considered here, the invasion speed is, roughly, a convex function of mean dispersal rates. Hence, fluctuations in mean dispersal distances increase invasion speeds. Intuitively, these fluctuations generate occasional bursts of long distance  dispersal events and it is these occasional long distance  dispersal events that effectively determine the invasion speed. This is consistent with the observation that dispersal kernels with ``fatter tails'' (e.g. Laplacian versus Gaussian) yield faster invasion 
speeds ~\citep{kot-etal-96}. Extending our analysis to reaction-diffusion models may explain a similar phenomena observed by \citet{seo-lutscher-11}. For populations with a sedentary reproductive stage and a randomly diffusing non-reproductive stage, \citet{seo-lutscher-11} found that periodic fluctuations in diffusion rates also increase invasion speeds.

Not all forms of temporal variation in dispersal enhance invasion speeds. Populations may exhibit multiple modes of dispersal either due to multiple dispersal vectors (polychory)~\citep{berg-83,bullock-etal-06,nathan-etal-08}, multiple behavioral types within the population~\citep{fryxell-etal-08,cote-etal-11}, or different passive forms of transport~\citep{hampe-04}.  The mean dispersal distances traveled in these modes can be substantially different. For example, \citet{jordano-etal-07} found that 50\% of small passerines dispersed seeds of St. Luce Cherry less than 51 meters, while 50\% of mammals dispersed seeds more than 495 meters. Our sensitivity analysis reveals that temporal fluctuations in the relative frequencies of these modes of dispersal, in of themselves, reduce invasion speeds.

The negative impact of fluctuations in the frequencies of dispersal modes or fluctuations in local demography can be  reversed when these fluctuations are positively correlated  with fluctuations in the ``appropriate'' mean dispersal distances. For example, if an increase in  the frequency of a particular dispersal mode correlates with longer dispersal distances traveled by individuals in this mode, then variation in the frequencies of this dispersal mode can increase invasion speeds. Similarly, when variation in the local demography of individuals in a particular state positively correlates with the mean dispersal distance traveled by these individuals (e.g. fecundity covarying with natal dispersal distances), this variation increases invasion speeds. Hence, unlike correlations between local demographic rates which can only dampen the negative effects of environmental fluctuations on population growth~\citep{tuljapurkar-90} and invasion speed, correlations between dispersal and demography can reverse the negative effects of fluctuations in local demography on invasion speed. We hypothesize that correlation between dispersal and local demography will be the rule rather than the exception. For example, larger plants will tend to produce more seeds, and release more seeds at a greater height so that they will tend to disperse further. Conditions allowing above-average growth for a plant species are therefore likely to result in higher fecundity and greater long-distance dispersal. 

For periodically-forced reaction diffusion equations, \citet{seo-lutscher-11} numerically demonstrated similar results for populations where individuals are born into a  mobile, non-reproductive stage and mature into a sedentary, reproductive stage. They found that periodic fluctuations in birth rates could increase invasion speeds provided that these fluctuations were positively correlated with fluctuating diffusion rates.  Interestingly, when individuals continuously move back and forth between mobile stages and reproductive stages, \citet{seo-lutscher-11} found that negative correlations, rather than positive correlations, between birth rates and diffusion rates increase invasion speeds. 
 
Our results relate to the seminal work of \citet{berg-83} (cf.~\citet{nathan-etal-08}) who emphasized the importance of `chance dispersal' for long distance dispersal. Berg argued that chance dispersal comes in two differing forms. The first form involves ``an unusually favorable combination of the regular dispersal factors'', occasions when many things happen to go well for dispersal at the same time. We have shown that positive correlations of dispersal distance with local demography or with frequencies of dispersal modes are ``unusually favorable combinations'' of factors that promote population spread. According to Berg, the second form of 'chance dispersal' corresponds to ``an unusual coincidence involving a dispersal factor not normally operating together with the taxon in question.'' In our context, this ``unusual coincidence'' would represent a rare mode of long-distance dispersal. Our analysis reveals that even infrequent occurrences of  a model leading to especially far-distant dispersal can produce a discontinuous and potentially large increase in invasion speed. This discontinuity has also been observed in temporally homogeneous models where a small fraction of the population continuously uses a mode of long distance dispersal. For example, \citet{neubert-caswell-00} estimated spatial spread rates for a neotropical plant \emph{Calathea ovandensis} whose seeds are dispersed by four ant species~\cite{horvitz-schemske-86}. They found that the invasion speed is mainly determined by one ant species which disperses only 7\% of the seeds but disperses the seeds the longest distance.
 
We illustrated the synergistic effects of positively correlated fluctuations in dispersal distances and fecundity with a size-structured patch model of perennial pepperweed, an invasive crucifer of the western United States. When temporal fluctuations are sufficiently large and positive correlated with fecundity (e.g. correlated with preciptation), these correlated fluctuations increased invasion speeds by approximately 60\% (cf. Fig.~\ref{fig:Pepperweed1}). Our sensitivity analysis of the pepperweed model also suggests that mortality of established pepperweed patches presents by far the best target for management. Even in the scenario that maximizes
pepperweed spread (large, positively correlated variability in dispersal and demography),
reducing the annual survivorship of established patches to 60\% would push the population into
decline and retreat $(\lambda_S \approx 0.8)$. Perennial pepperweed can be controlled 
by herbicide treatments, which achieve nearly 100\% patch eradication \citep{hutchinson-viers-2011}, but
non-chemical means are preferable or necessary in some settings (e.g., organic agriculture or natural areas). 
The results from our model support the conclusion of \cite{hutchinson-viers-2011} that 
some non-chemical methods would also be effective enough,
such as a mow, till and tarp treatment that was found to eradicate 50\% of patches completely, and reduce
by 95\% or more an additional 25\% of patches.
 
Collectively, our results illustrate the importance of fluctuations in dispersal kernels for rates of population spread. However, currently there are very few empirical studies which have measured these fluctuations, let alone their correlation with local demography. While estimating dispersal kernels is notoriously difficult, our work highlights that understanding a single snap shot in time of dispersal kernels may not be adequate for estimating invasion speeds. In particular, ignoring temporal variation in dispersal rates can substantially underestimate rates of spatial spread. From a theoretical perspective, our analysis raises the issue of whether temporal variation in patterns of dispersal alters other aspects of ecological dynamics, such as persistence or coexistence in spatially heterogeneous environments. Dispersal variation may also be important for processes other than species invasions, such as the spread of alleles (either new natural mutations or alleles escaping from a GMO crop) and the spread of novel pathogens. Because patterns of spread in alleles and rapidly-evolving pathogens such as flu can often be reconstructed through genetic analyses (e.g., \citet{bedford-2010}), the data needed for parameterizing models and testing predictions about spread rates might be most readily available for these applications. 
 
\bigskip 

\textbf{Acknowledgments} We are grateful to Joshua Viers, Rachel Hutchinson, Jim Quinn and Claire Stouthamer for advising us
about pepperweed biology and modeling. For comments on the manuscript we thank Hal Caswell, Ben Dalziel, Matt Holden, Rachel
Hutchinson, Katherine Marchetto, and Peter Ralph. SPE's work was partially supported by the Institute for Computational Sustainability at Cornell, which is supported by an NSF Expeditions in Computing grant (award 0832782) and by the 
Atkinson Center for a Sustainable Future at Cornell University. SJS's work was partially supported by funds 
from the Dean's Office of the College of Biological Sciences at University of California, Davis.

%\bibliographystyle{elsarticle-harv}
%\bibliography{StochWaveTPB3}

\appendix

\centerline{\textbf{\Large{Appendices}}}

\section{Calculating the wave speed}
\label{Appendix:WaveSpeed}
Let $\gamma(s)$ denote the dominant Lyapunov exponent for $H_{s,t}$.
To find possible traveling waves, we consider population states of the form
\begin{equation}
n_t(x,z)=u_t(z)e^{-sx}.
\label{eqn:WaveShape}
\end{equation}
Assuming that this form holds at time $t$, we have (using the spatial invariance and symmetry of the kernel)
\begin{equation}
\begin{aligned}
n_{t + 1}(x,z) & = \iint K_t(x - x_1,z,z_1)u(z_1,t) e^{ - s{x_1}} d x_1 d z_1 \\
& = e^{-sx} \iint K_t(x - x_1,z,z_1)u(z_1,t) e^{- s(x_1-x)} d x_1 d z_1 \\
& = e^{-sx} \iint K_t(v,z,z_1)u(z_1,t)e^{- sv} d x_1 d z_1 \\
& = e^{-sx} \int H_{s,t}(z,z_1)u(z_1,t) d z_1.
\end{aligned}
\end{equation}
Therefore, if equation \eqref{eqn:WaveShape} holds initially, then by induction it holds for all time with
$u_{t+1}=H_{s,t}u_t.$
The limit to the area occupied by the population at time $t$, $X_t(s)$, is defined by the property that
$$n_c = \left\langle {{n_t}({X_t}(s)),\bw} \right\rangle  = e^{-sX_t(s)}\left\langle{ \bu_t, \bw} \right\rangle $$ 
where $\bw(z)$ is a non-negative weighting function (e.g., the biomass of a state-$z$ individual).  
Then 
\begin{equation}
X_t(s) = \frac{1}{s} \log \left\langle{ u_t, \bw} \right\rangle - \log n_c.
\label{eqn:Xts}
\end{equation}
and therefore $$\frac{X_t(s)}{t} \to \frac{\gamma(s)}{s} \mbox{ as } t \to \infty$$
with probability 1. In addition, if the environment process is uniformly mixing, then
$t^{-1} \log \left\langle{ u_t, \bw} \right\rangle$ has an asymptotic normal distribution with mean
$\gamma(s)$ and variance $\sigma_2/t$ for some $\sigma$. The average rate of spread up to finite time $t$, given
by $X_t(s)/t$, is therefore asymptotically Normal with mean $\gamma(s)$ and variance decreasing in proportion
to $t^{-1}$.

A standard monotonicity argument (see \cite{kot-etal-96} or Appendix A in \cite{schreiber-ryan-11})
shows that any model solution with the initial population limited to a finite spatial domain cannot 
spread faster than the slowest such wave speed, given by equation \eqref{eqn:cstar}.   

\section{Sensitivity analysis of the spread rate}\label{Appendix:sensitivity}
Because $c^*$ is defined implicitly, for sensitivity analysis of $c^*$ we need to use the
following general result. Suppose $F(\theta) = \mathop {\min }\limits_s f(s,\theta )$
where $s,\theta$ are real and and $f$ is a smooth real-valued function. Let $s^*(\theta)$ denote the
value of $s$ at which $f(\cdot,\theta)$ is minimized. 
Then $\frac{\partial f}{\partial s}(s^*(\theta),\theta) = 0$
and
\be
\frac{\partial F}{\partial \theta } = \frac{\partial}{\partial \theta }f({s^*(\theta)},\theta )  
= \left[ {\frac{\partial f}{\partial s}\frac{{\partial {s^*}}}{{\partial \theta }}
+ \frac{{\partial f}}{{\partial \theta }}} \right]({s^*(\theta) },\theta )
= \frac{{\partial f}}{{\partial \theta }}({s^*(\theta) },\theta ).
\label{eqn:sensF}
\ee
Applying this to equation \eqref{eqn:cstar}, we have
\be
\frac{\partial c^*}{\partial \theta} = \frac{1}{s^*}\frac{\partial \gamma (s^*)}{\partial \theta}.
\label{eqn:csens}
\ee
We can take $\theta$ to be a kernel entry $K_t(v,y,x)$, an entry in one of the component kernels
$F_t, P_t, k_{d,t}$, or an underlying parameter in one of the kernels (e.g., the slope
parameter in a regression model for survival probability as a function of size). A small perturbation
to any of these has the effect of perturbing the transformed kernel $H_{s,t}$ to $H_{s,t} + \varepsilon \bC_t$
for $\varepsilon \ll 1$.

There are two basic formulas for the resulting change in $\gamma$ (\cite{rees-ellner-09}; 
note that \cite{rees-ellner-09} refer to $\gamma$ as $\log \lambda_S$).
The first, which applies to an arbitrary perturbation kernel $\bC_t$, is
\be
\frac{\partial \gamma}{\partial \varepsilon } =
\E\left[ {\frac{{\left\langle {{v_{s,t + 1}},{C_t}{w_{s,t}}} \right\rangle }}
{{\left\langle {{v_{s,t + 1}},{H_{s,t}}{w_{s,t}}} \right\rangle }}} \right]
\label{eqn:pertform1}
\ee
where $v_{s,t},w_{s,t}$ are time-dependent stationary reproductive value and population structure
sequences of the base kernel $H_{s,t}$. The second concerns the situation where the
right-hand size of \eqref{eqn:pertform1} is zero, because the perturbation kernel has
zero mean and is independent of the $H_{s,t}$. In that case, the change in $\gamma$ is $O(\epsilon^2)$
and to leading order
\be
\gamma(s,\varepsilon)=  
\gamma(s,0)- \frac{\varepsilon^2}{2}\mathrm{Var}\left[ {\frac{{\left\langle {{v_{s,t + 1}},{C_t}{w_{s,t}}} \right\rangle }}
{{\left\langle {{v_{s,t + 1}},{H_{s,t}}{w_{s,t}}} \right\rangle }}} \right] =
\gamma(s,0) - \frac{\varepsilon^2}{2}\E\left[ {\frac{{{{\left\langle {{v_{s,t + 1}},{C_t}{w_{s,t}}} \right\rangle }^2}}}
{{{{\left\langle {{v_{s,t + 1}},{H_{s,t}}{w_{s,t}}} \right\rangle }^2}}}} \right].
\label{eqn:pertform2}
\ee
Applying \eqref{eqn:pertform2} to perturbations consisting of small fluctuations around a constant mean kernel, we
get the small-variance approximation,
\be
\gamma \approx \log {\lambda_0} - \frac{{\mathrm{Var}\left\langle {v_s, H_{s,t}w_s} \right\rangle }}{{2\lambda_0^2}}
\label{eqn:SmallVar}
\ee
where $\lambda_0$ is the dominant eigenvalue of the mean kernel $\E[H_{s,t}]$ and $v_s,w_s$ are the dominant left
and right eigenfunctions of the mean kernel normalized so that $\left\langle v_s,w_s \right \rangle = 1$.
\cite{rees-ellner-09} explain how these formulas can be used to compute the various sensitivities and elasticities that
have been proposed for stochastic matrix and integral models, and the sensitivity to perturbations of the vital rate
functions that are used to construct the kernel.

Some of our results depend on the fact that \eqref{eqn:pertform2} also holds under
weaker assumptions than those stated above. In particular, the calculations
in \cite{rees-ellner-09} shows that \eqref{eqn:pertform2} holds so long as the conditional mean of
any perturbation kernel $C_t$ given everything else (the full sequence of unperturbed kernels and
of all other perturbation kernels) is identically zero. For example, \eqref{eqn:pertform2} applies to  
perturbation kernels $z_t F_t$ where the $z_t$ are \textit{iid} random scalars with zero mean and independent
of the unperturbed kernels, even though this perturbation is not independent of the unperturbed kernel process.  

\section{Moment generating functions of dispersal kernels}
\label{Appendix:mgf}
We collect here some elementary but useful properties related to the moment generating functions
of dispersal kernels. Let $\mu(dx)$ denote a probability distribution for
spatially homogeneous dispersal with displacement $x$, meaning that $\mu$ is a measure on the real line and
$\mu(A)=\mu(-A)$. Let $X$ denote a random variable
with distribution $\mu$ and $M(s)=\E[e^{sX}]$ its moment generating function. We assume that $M(s)$ is
finite on some (necessarily symmetric) open interval $(-s_1,s_1)$.

\begin{enumerate}
\item $M(s)$ is infinitely differentiable on $(-s_1,s_1)$ and its $n^{th}$ derivative is given by
\be
M^{(n)}(s) =\E[X^n e^{sX}].
\label{eqn:Mn}
\ee
\textit{Proof:} This seems to be widely known but we have not found a proof in print.
For $s=0$ this is a standard result about the moment generating function. The symmetry of $\mu$
implies that $M(s)=M(-s)$ so it suffices to consider $s>0$.
We proceed by induction on $n$ for an arbitrary fixed $s \in (0,s_1)$.  For $n=1,  M'(s)$ is the limit
as $\varepsilon \to 0$ of
\be
\E\left[ {\frac{{{e^{(s + \varepsilon )X}} - {e^{sX}}}}{\varepsilon }} \right]
\label{eqn:Mprime}
\ee
The integrand  in \eqref{eqn:Mprime} converges pointwise to $Xe^{sX}$, as needed. To prove convergence of the integral we consider
$X\ge 0$ and $X<0$ separately, and find finitely $\mu$-integrable upper bounds on the absolute value of the integrand,
so that the Dominated Convergence Theorem applies. By Taylor's Theorem, the integrand in \eqref{eqn:Mprime}
is pointwise equal to $Xe^{s^*X}$ for some $s^*$ (depending on $X$) between $s$ and $s+\varepsilon$.
For $X \ge 0$, pick $\sigma_1 \in (s,s_1)$. Then for for all $\varepsilon$ sufficiently small and $X \ge 0$,
$Xe^{s^* X} \le Xe^{\sigma_1 X} = Xe^{-\delta X}e^{(\sigma_1+\delta) X}.$
Choose $\delta$ so that $\sigma_1 + \delta < s_1$.
For $X \ge 0, Xe^{-\delta X}$ is positive and no larger than  $1/(\delta e).$ So for $X\ge 0$ the
integrand in \eqref{eqn:Mprime} is bounded above by a constant multiple of $e^{(\sigma_1+\delta) X}$,
which is integrable because $\sigma_1 + \delta < s_1$. For $X<0$, the integrand is bounded above
in absolute value by $1/(s^* e)$ which is uniformly bounded for $\varepsilon$ small.   

Assuming the result for $n$, the difference quotient for $M^{(n+1)}$ is
$\E \left[ \left(X^n{{e^{(s + \varepsilon )X}} - X^n{e^{sX}}}\right)/\varepsilon \right].$
The construction of upper bounds for the integrand is nearly identical to the case $n=1$, using
the fact that $x^{n+1} e^{-\delta x}$ for $\delta >0, x>0$ has a finite maximum.  

\item Unless $\mathrm{Var}[X]=0$, $M^{(n)}(s)$ is positive for $n$ even, and has the sign of $s$ for $n$ odd.
\textit{Proof:} For $n$ even this follows immediately from \eqref{eqn:Mn}. For $n$ odd, because $X$ and
$-X$ are identically distributed, we have
$ 2M^{(n)}(s) =\E[X^n e^{sX} + (-X)^n e^{-sX}] = \E[X^n( e^{sX} - e^{-sX})]$, and the integrand has the sign
of $s$ pointwise except at $X=0$.

\item $\gamma(s)$ is nondecreasing on $(0,s_1).$ \textit{Proof:} The kernel $K_t(v,z,z_0)$ can be factored
as $\tilde K_t(z,z_0)\mu_t(dv|z,z_0)$ where $\mu_t(dv|z,z_0)$ is the probability measure for
state-$z$ individuals at time $t+1$ produced by state-$z_0$ individuals at time $t$. The last
result, applied to the moment generating functions of the $\mu_t$, implies that the transformed
kernels $H_{s,t}$ are monotonically increasing in $s$ on $(0,s_1)$, therefore $\gamma(s)$ must
be everywhere nondecreasing. If the variability is temporally independent, then equation
\eqref{eqn:pertform1} implies that $\gamma(s)$ is strictly increasing.  

\item $M^{(2)}M > (M^{(1)})^2$ unless $\mathrm{Var}[X]=0$.
\textit{Proof:} using equation \eqref{eqn:Mn}, this is the
Cauchy-Schwartz inequality applied to the functions $e^{sX/2},Xe^{sX/2}$, with strict inequality
because the two functions are not proportional to each other.

\item Equation \eqref{eq:modes2} is always positive. \textit{Proof:}
We will show that 
\begin{equation}\label{aa}
2\sum_i \bar p_i m_i (\bar L_i s^*)\left( \sum_i \bar p_i m_i''(\bar L_i s^*) (s^*)^2 \mbox{Var}[\ell_i(t)]/2 - \frac{\langle v, F w \rangle}{2\lambda_0} \mbox{Var}\left[ \sum_i \bar p_i m_i'(\bar L_i s^*) s^* \ell_i(t)\right]\right)\ge 0
\end{equation} 
with a strict inequality if $\mbox{Var}[\ell_i(t)]>0$ for some $i$. 
To simplify the notation, define 
\[
a_i^2=\bar p_i m_i (\bar L_i s^*) \quad b_i^2 = \bar p_i m_i''(\bar L_i s^*) (s^*)^2 \mbox{Var}[\ell_i(t)] \quad c_i =\bar p_i m_i'(\bar L_i s^*) \sqrt{\mbox{Var}[\ell_i(t)]} s^*.
\]
Using this notation and noting that $\mbox{Cov}[ \ell_i(t),\ell_j(t)] \le \sqrt{\mbox{Var}[\ell_i(t)]\mbox{Var}[\ell_j(t)]}$, we get that the left hand side of \eqref{aa} is bounded below by 
\begin{equation}\label{bb}
\sum_i a_i^2 \left( \sum_i b_i^2 - \frac{\langle v, F w \rangle}{\lambda_0} \left(\sum_i c_i\right)^2 \right).
\end{equation} 
Using the fact that $\lambda_0 = \langle v, H_{s^*}^0 w \rangle \ge \langle v, \sum_i \bar p_i m_i(\bar L_i s^*) F w\rangle=\sum_i a_i^2 \langle v, F w \rangle$, we get that \eqref{bb} is bounded below by 
\begin{equation}\label{cc}
\sum_i a_i^2 \left( \sum_i b_i^2 - \frac{1}{\sum_i a_i^2} \left(\sum_i c_i\right)^2 \right)=\sum_i a_i^2  \sum_i b_i^2 -  \left(\sum_i c_i\right)^2 .
\end{equation} 
Because $m_i''(\bar L_i s^*) m_i (\bar L_i s^*) > (m_i'(\bar L_i s^*)^2$ as proved above we have that $a_i b_i \ge c_i$ with 
strict inequality when $\mbox{Var}[\ell_i(t)] >0$.  This observation plus the Cauchy-Schwartz inequality imply that 
\[
\sum_i a_i^2  \sum_i b_i^2 -  \left(\sum_i c_i\right)^2 \ge \left( \sum_i a_i b_i \right)^2 - \left(\sum_i c_i \right)^2 \ge 0 
\]
with a strict inequality if $\mbox{Var}[\ell_i(t)]>0$ for some $i$. 
\end{enumerate}

\section{Two-parameter perturbations}
\label{TwoParameter}
To understand the effect of the random perturbation \eqref{eqn:SmallVarM} on the invasion speed, we initially consider the two parameter perturbation
$H_{s,t}+\varepsilon_1 C_{s,t}+\varepsilon_2 D_{s,t}$ where $H_{s,t}=P+M_1(\bar Ls) F$, $C_{s,t}=s (L_t-\bar L) M_1'(\bar Ls)F$, and $D_{s,t}=s^2(L_t-\bar L)^2M_1''(\bar Ls)F/2$. By Taylor's theorem, we have the second order approximation
\[
\gamma(\varepsilon_1,\varepsilon_2,s) \approx \gamma(0,0,s)+ \frac{\partial \gamma}{\partial \varepsilon_1}(0,0,s) \varepsilon_1+ \frac{\partial \gamma}{\partial \varepsilon_2}(0,0,s) \varepsilon_2+  \frac{1}{2}\frac{\partial^2 \gamma}{\partial \varepsilon_1^2}(0,0,s) \varepsilon_1^2+ \frac{1}{2}\frac{\partial^2 \gamma}{\partial \varepsilon_2^2}(0,0,s) \varepsilon_2^2+ \frac{\partial ^2 \gamma}{\partial \varepsilon_1 \partial \varepsilon_2}(0,0,s) \varepsilon_1\varepsilon_2
\]
By sensitivity formula \eqref{eqn:s1}, we have
\[
\frac{\partial \gamma}{\partial \varepsilon_2}(0,0,s) =\frac{s^2 \sigma_L^2 M_1''(\bar L s) \left\langle v_s,Fw_s \right\rangle}{2 \lambda_0(s)}
\]
where $\sigma_L^2=\mathrm{Var}[L_t]$ and $v_s,w_s$ the corresponding left and right eigenfunctions of $H_{s}$ scaled so that
$\left \langle v_s,w_s \right\rangle =1$.
By sensitivity formula \eqref{eq:s2}, we have
\[
\frac{\partial \gamma}{\partial \varepsilon_1}(0,0,s)=0 \mbox{ and } \frac{\partial^2 \gamma}{\partial \varepsilon_1^2}(0,0,s)=-\frac{s^2 \sigma_L^2 (M_1'(\bar L s))^2 \left\langle v_s,Fw_s \right\rangle^2}{\lambda_0(s)^2}.
\]
Setting $\varepsilon_1=\varepsilon$, $\varepsilon_2=\varepsilon^2$, and only considering terms up to order $\varepsilon^2$, we get
\begin{eqnarray*}
\gamma(\varepsilon_1,\varepsilon_2,s) &\approx &\gamma(0,0,s)+0+\frac{s^2 \sigma_L^2 M_1''(\bar L s) \left\langle v_s,Fw_s \right\rangle}{2 \lambda_0(s)} \varepsilon^2 -\frac{s^2 \sigma_L^2 (M_1'(\bar L s))^2 \left\langle v_s,Fw_s \right\rangle^2}{2\lambda_0(s)^2}\varepsilon^2\\
&=&\gamma(0,0,s)+\frac{\varepsilon^2 s^2\sigma_L^2 \left\langle v_s,Fw_s \right\rangle}{2\lambda_0(s)}\left(M_1''(\bar L s) -\frac{ (M_1'(\bar L s))^2 \left\langle v_s,Fw_s \right\rangle}{\lambda_0(s)}\right)
\end{eqnarray*}

\section{Sensitivity calculations for pepperweed}
We explain here how the general perturbation formulas were used to do the calculations for
the pepperweed model. The elasticity of $\lambda_S$ to the mean of a size-specific
fecundity $\bar F(y,x)$ is by definition 
$$e(\lambda_S,\bar F(y,x)) = \frac{\bar F(y,x)}{\lambda_S}\frac{\partial \lambda_S}{\partial \bar F(y,x)}
= \frac{\bar F(y,x)}{\lambda_S}\frac{\partial \lambda_S}{\partial \bar K(y,x)}$$ 
the last equality holding because $K=P+F$. From Table 3 in \cite{rees-ellner-09} we have
$$ \frac{\partial \lambda_S}{\partial \bar K(y,x)} = s_S(y,x) 
= \lambda_S \E\left[\frac{v_{t+1}(y)w_t(x)}{\langle v_{t+1},K_t w_t \rangle} \right].$$
We therefore have 
\be
e(\lambda_S,\bar F(y,x)) = \bar F(y,x) \E\left[\frac{v_{t+1}(y)w_t(x)}{\langle v_{t+1},K_t w_t \rangle} \right].
\ee
In parallel, 
\be
e(\lambda_S,\bar P(y,x)) = \bar P(y,x) \E\left[\frac{v_{t+1}(y)w_t(x)}{\langle v_{t+1},K_t w_t \rangle} \right].
\ee

The elasticity of $c^*$ to the mean of a size-specific fecundity $\bar F(y,x)$ is by definition 
\be
e(c^*,\bar F(y,x)) = \frac{\bar F(y,x)}{c^*}\frac{\partial c^*}{\partial \bar F(y,x)}.
\ee
Because $H_{s,t}=M_t(s)F_t + P_t$, a unit perturbation in $F_t(y,x)$ increases $H_{s,t}(y,x)$
by $M_t(s)$. The corresponding perturbation kernel $C_t$ in \eqref{eqn:s1} is therefore 
$M_t(s)\delta_{y,x}$ where $\delta_{y,x}$ is an approximate delta-function centered at $(y,x)$, and
the sensitivity at $(y,x)$ is the limiting value as the support of $\delta_{y,x}$ shrinks to
a point (\cite{ellner-rees-06}). We therefore have 
\be
\frac{\partial c^*}{\partial \bar F(y,x)}
= \frac{1}{s^*}\E\left[\frac{M_t(s^*)v_{s^*,t+1}(y)w_{s^*,t}(x)}{\langle v_{s^*,t+1},H_{s^*,t} w_{s^*,t} \rangle} \right].
\ee
Because $c^*=\gamma(s^*)/s^*$, we get
\be
e(c^*,\bar F(y,x))
=\frac{\bar F(y,x)}{\gamma(s^*)}\E\left[\frac{M_t(s^*)v_{s^*,t+1}(y)w_{s^*,t}(x)}{\langle v_{s^*,t+1},H_{s^*,t} w_{s^*,t} \rangle} \right].
\ee
For a perturbation to size-specific survival the factor $M_t(s)$ is absent, so we get 
\be
e(c^*,\bar P(y,x))
=\frac{\bar P(y,x)}{\gamma(s^*)}\E\left[\frac{v_{s^*,t+1}(y)w_{s^*,t}(x)}{\langle v_{s^*,t+1},H_{s^*,t} w_{s^*,t} \rangle} \right].
\ee

The expectations in the formula above can be computed for all $(y,x)$ from one long simulation of the model using 
the methods described by \cite{rees-ellner-09}. For an IPM implemented using midpoint rule, these are 
the same as the standard methods for stochastic matrix models. For all such calculations we used 10000-year 
simulations, omitting ``burn-in'' periods of 250 years at the beginning and the end to allow convergence 
of the $v$ and $w$ vectors to their stationary distributions.  

For sensitivities to an underlying parameter such as the mean patch growth rate, the simplest approach is brute force: 
perturb the parameter by $\pm \varepsilon$, recompute wave speeds (reducing Monte Carlo error by using the same sequence 
of year types for all values of the parameter), and estimate the derivative by finite difference. 

\section{State-dependent dispersal}
\label{StateDependent}
We consider here the plant model \eqref{eqn:PlantKernel} with dispersal dependent on the
states of the parents and offspring, and small temporal variation in mean dispersal distance. Specifically,
we assume 
\be
k_{d,t}(v,z,z_0)= \frac{1}{L_t(z,z_0)}k(v/L_t(z,z_0); z,z_0)
\label{eqn:StateDepKernel}
\ee
with $(L_t-\bar L)$ uniformly small and constant in sign as a function of $(z,z_0)$ for each $t$. 
The temporal variability in $L_t$ is assumed to
be \textit{iid} and independent of the demographic variation. Note that the form of the dispersal
kernel can also depend on parent and offspring states. Our goal is to show that the effect of the  
(small) variance in $L_t$ is always to increase $\gamma$ and therefore to increase wave speed. 

We need the following general result. Let $f,g,h,w$ be bounded nonnegative functions 
on a finite measure space $(X,\mu)$ such that $fg \ge h^2$ with strict inequality on a set
of positive measure in the support of $w$. Then letting
$\langle \:, \;\rangle$ denote the $L_2$ inner product  on $(X,\mu)$  
\be
\label{eqn:yacsi}
\begin{aligned}
\left( \int_X{hw d\mu}\right)^2  & < \left( \int_X{\sqrt{fg}w d\mu}\right)^2 = \left( \int_X{\sqrt{fw} \sqrt{gw}d\mu}\right)^2 \\
& = \langle \sqrt{fw},\sqrt{gw} \rangle^2 
\le \langle \sqrt{fw},\sqrt{fw} \rangle  \langle \sqrt{gw},\sqrt{gw} \rangle \\
& = \left( \int_X{fw d\mu}\right)\left( \int_X{gw d\mu} \right). 
\end{aligned}
\end{equation}

The small-variance approximation to the mgf of the dispersal kernel \eqref{eqn:StateDepKernel} is 
\be
M(L_t s) = M(\bar L s + (L_t-\bar L)s) \approx 
M(\bar L s) + s M_1'(\bar L s)(L_t-\bar L)
+ \frac{s^2}{2} M''(\bar L s)(L_t-\bar L)^2
\label{eqn:SmallVarSDM}
\ee
with all terms in \eqref{eqn:SmallVarSDM} being functions of $(z,z_0)$. Let $\circ$ denote 
element-by-element multiplication, i.e. $(f \circ g)(z,z_0) = f(z,z_0)g(z,z_0)$. 
The expansion of the transformed kernels $H_{s,t}$ corresponding to \eqref{eqn:SmallVarSDM} is 
\be
H_{s,t} \approx  P_t +  M(\bar L s)\circ F_t + s M'(\bar L s)\circ (L_t - \bar L) \circ F_t 
+ \frac{s^2}{2} M''(\bar L s) \circ (L_t - \bar L)^2 \circ F_t.
\label{eqn:SmallVarSDH}
\ee
Because of the independence of $L_t$, the $M'$ term in \eqref{eqn:SmallVarSDH} has identically zero 
mean conditional on the unperturbed process. Its effect on $\gamma(s)$ is therefore to leading order (omitting the
$s$-dependence in $v,w$ and $M$ and its derivatives) 
\be
-\frac{s^2}{2}\E\left[
\frac{\left\langle v_{t+1}, ((L_t - \bar L) \circ M' \circ F_t) w_t \right\rangle^2}
{ \left\langle v_{t+1}, H^0_t w_t \right\rangle^2} \right].
\label{eqn:M1term}
\ee
where $H^0 = P_t +  M(\bar L s)\circ F_t$ is the unperturbed kernel. The $M''$ term has nonzero mean, 
so its leading order effect on $\gamma$ is 
\be
\frac{s^2}{2}\E\left[
\frac{\left\langle v_{t+1}, ((L_t - \bar L)^2 \circ M'' \circ F_t) w_t \right\rangle}
{ \left\langle v_{t+1}, H^0_t w_t \right\rangle} \right].
\label{eqn:M2term}
\ee
We aim to show that \eqref{eqn:M2term} is larger than \eqref{eqn:M1term}. We do this by showing
that the integrand in \eqref{eqn:M2term} is with probability 1 larger than the integrand in
\eqref{eqn:M1term}, i.e.,    
\be
\langle v_{t+1}, ((L_t - \bar L)^2 \circ M'' \circ F_t) w_t \rangle 
\langle v_{t+1}, H^0_t w_t \rangle > 
\langle v_{t+1}, ((L_t - \bar L) \circ M' \circ F_t) w_t \rangle^2.
\ee
Neither side of the inequality is affected if we replace $L_t - \bar L$ with its
absolute value, so suffices to consider $L_t(z,z_0) -\bar L > 0 $ (recall that $L_t-\bar L$ is
assumed to have constant sign in any given year). Because $H^0 > M \circ F_t$ and $v_{t+1},w_t$ 
are both everywhere positive, it suffices to show that 
\be
\langle v_{t+1}, ((L_t - \bar L)^2 \circ M'' \circ F_t) w_t \rangle 
\langle v_{t+1}, (M \circ F_t) w_t\rangle \ge 
\langle v_{t+1}, ((L_t - \bar L) \circ M' \circ F_t) w_t \rangle^2.
\label{eqn:Mterms}
\ee
Writing out the inner products as integrals, equation \eqref{eqn:Mterms} follows from 
\eqref{eqn:yacsi} with $f=(L_t-\bar L)^2 \circ M'', g=M, h=(L_t-\bar L) \circ M'$ and 
$w(z,z_0)=v_{t+1}(z)F_t(z,z_0)w_t(z_0)$. 

\end{document}